\documentclass[final]{IEEEtran}

\usepackage{ifthen}
\newcommand{\PaperORReport}{Report}
\newcommand{\whichexample}{Localization}

\usepackage{amsmath,amsfonts,amssymb} 
\usepackage{mathrsfs}
\usepackage{bbm}
\usepackage{pstricks}
\usepackage{verbatim}
\usepackage{enumerate}

\usepackage{url} 
\usepackage{graphicx,psfrag}
\usepackage[center]{subfigure}

\makeatletter\let\NAT@parse\undefined\makeatother 

\usepackage[numbers,sort&compress]{natbib}
\usepackage[ieee,fancythm]{jphmacros2e}

\DeclareMathAlphabet\EuScript{U}{eus}{m}{n}
\DeclareFontFamily{OT1}{pzc}{}
\DeclareFontShape{OT1}{pzc}{m}{it}{<-> s * [1.200] pzcmi7t}{}
\DeclareMathAlphabet{\mathpzc}{OT1}{pzc}{m}{it}
\DeclareMathOperator{\Exp}{E}    
           
\newcommand{\G}{\mathcal{G}}
\newcommand{\V}{{\mathpzc{V}}}
\renewcommand{\E}{{\mathpzc{E}}} 

\newcommand{\R}{\mathbb{R}}

\newcommand{\N}{\mathbb{N}}                 
\newcommand{\Z}{\mathbf{Z}}

\newcommand{\scr}{\mathcal}

\renewcommand{\S}{\mathbb{S}}

\renewcommand{\i}{\mbf{i}}       
\newcommand{\volt}{V}

\renewcommand{\comment}[1]{}

\renewcommand{\ball}{\mathcal{B}}

\title{\LARGE \sc Error Scaling Laws for Linear Optimal Estimation from Relative
Measurements}

\author{Prabir~Barooah,~\IEEEmembership{Member,~IEEE,}
        Jo\~{a}o P. Hespanha,~\IEEEmembership{Fellow, IEEE}%
\thanks{Prabir Barooah is with the Dept.~of Mech.~and Aero.~Engg., Univ.~of Florida, Gainsville, FL 32611 (email: pbarooah@ufl.edu)}
\thanks{Jo\~{a}o P. Hespanha is with the Center for Control, Dynamical-systems, and Computation, and the Dept.~of Elec.~and Comp.~Engg., Univ.~of California, Santa Barbara, CA 93106 (email: hespanha@ece.ucsb.edu).}
\thanks{This material is based upon work supported by the Institute for
Collaborative Biotechnologies through grant DAAD19-03-D-0004 from the U.S.
Army Research Office.}}

\begin{document}

\maketitle

\begin{abstract}
  We study the problem of estimating vector-valued variables from
  noisy ``relative'' measurements. This problem arises in several
  sensor network applications. The measurement model can be expressed
  in terms of a graph, whose nodes correspond to the variables and
  edges to noisy measurements of the difference between two
  variables. We take  an arbitrary variable as the reference and consider the optimal (minimum variance) linear unbiased estimate of the remaining variables.

 We investigate how the error in the optimal linear unbiased estimate of a node
variable grows with the distance of the node to the reference
node. We establish a classification of graphs, namely, dense or sparse
in $\R^d,\;1\leq d \leq 3$, that determines how the linear unbiased optimal estimation 
error of a node grows with its distance from the reference node. In
particular, if a graph is dense in $1$,$2$, or $3$D, then a node
variable's estimation error is upper bounded by a linear, logarithmic,
or bounded function of distance from the reference, respectively. Corresponding
lower bounds are obtained if the graph is sparse in $1$, $2$ and $3$D.
Our results also show that naive measures of graph
density, such as node degree, are inadequate predictors of the
estimation error. Being true for the optimal linear unbiased estimate, these scaling laws determine algorithm-independent limits on the estimation accuracy achievable in large graphs. 
\end{abstract}


\section{Introduction}\label{sec:intro}
\ifthenelse{\equal{\whichexample}{TimeSync}}{
Several applications in sensor and actuator networks lead to
estimation problems where a number of variables are to be estimated
from noisy measurements of the difference between certain pairs of
them.  Consider the problem of synchronizing the local times at the nodes of a network. The  local time at node $u$ when the time at a reference clock (that is fixed arbitrarily at one of the clocks of the network) is $t$, is $t_u(t) = \alpha_u t + \beta$, where $\alpha_u$ is the \emph{skew} and $\beta_u$ is the \emph{offset} of the node $u$. None of the clocks know their skews and offsets, but it is possible for a pair of nodes $u$ and $v$ to measure the difference between their offsets $\beta_u - \beta_u$ and log-skews $log(\alpha_u) - log(\alpha_v)$ up to some error by exchanging time-stamped packets. Figure~\ref{fig:time-sync} describes how difference of offset measurements can be obtained when the skews are unity; see~\cite{Karp_03TimeSync,BarooahThesis:07} for a details on how these measurements can be obtained when skews are not unity. Both these situations are special cases of the following: a pair of nodes $u$ and $v$ has access to the measurement
\begin{align}\label{eq:measurement}
  \zeta_{u,v} = x_u - x_v + \epsilon_{u,v},
\end{align}
where $\epsilon_{u,v}$ denotes measurement error and $x_u,x_v \in \R^k$ ($k \geq 1$) are unknown variables. The problem of interest is to use the $\zeta_{u,v}$'s to estimate the variables $x_v$  with respect to a common reference frame whose origin is assigned arbitrarily to the variable at one of the nodes. 

A similar problem occurs in localization of sensor nodes,  where a sensor does not
know its position in a global coordinate system, but may be able to measure its
position relative to a set of nearby nodes. In that case, the variables $x_u$ are positions of the sensor nodes in 2- or 3-dimensional space, and the problem of
interest is to use the $\zeta_{u,v}$'s to estimate the positions of
all the nodes in a common coordinate system whose origin is fixed
arbitrarily at one of the nodes. 

\medskip

Such estimation problems arise not only in time-synchronization~\cite{Karp_03TimeSync,Kumar-timesync-II:06,Kumar-timesync-I:06} and localization~\cite{AB_JG_JM_GS_MobiHoc:06} but also in motion consensus in sensor-actuator networks~\cite{BarooahHespanhaDec06a}; see~\cite{PB_JPH_CSM:07,BarooahHespanhaDec06a} for an overview of these applications. Motivated by these applications, we study the
problem of estimating vector valued variables from noisy measurements
of the difference between them.  In particular,
denoting the variables of interest by $\{x_i:i\in\mathbf{V}\}$ where
$\mathbf{V}:=\{1,2,\dots\}$, we consider problems for which noisy
relative measurements of the form~\eqref{eq:measurement} are
available.  The ordered pairs of indices $(u,v)$ for which we have relative
measurements form a set $\mathbf{E}$ that is a (typically strict)
subset of the set $\V\times\V$ of all pairs of indices.  Just with
relative measurements, the $x_u$'s can be determined only up to an
additive constant.  To avoid this ambiguity, we assume that a
particular variable (say $x_o$) is used as the \emph{reference}, which is therefore assumed known. The problem of interest is to estimate the remaining node variables from all the available measurements.

\medskip

The measurement equations \eqref{eq:measurement} can be naturally
associated with a directed graph $\G=(\V,\E)$ with an edge from node
$u$ to $v$ if the measurement $\zeta_{u,v}$ is available.  The graph
$\G$ is called the \emph{measurement graph}, and each vector $x_u$,
$u\in\V$ is called the \emph{$u$-th node variable}. The measurement
noise $\epsilon_{e}, e\in\E$ is assumed zero mean and spatially
uncorrelated, i.e., $\Exp[\epsilon_e]=0 \; \forall e \in\E$ and
$\Exp[\epsilon_e \epsilon_{\bar e}^T] =0$ if $e \neq \bar{e}$.

\medskip

\begin{figure*}[t]
\psfrag{u}{$u$} \psfrag{v}{$v$} \psfrag{w}{$w$}
\centering \includegraphics[scale=0.4]{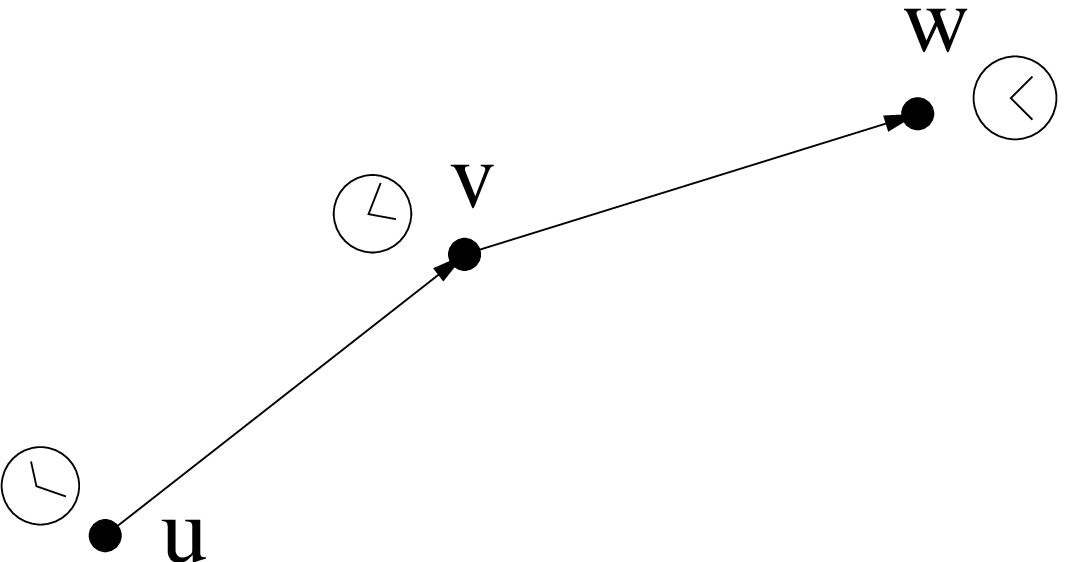}
\caption{Measurement of differences in
local times by exchanging time-stamped messages. Node $u$ transmits a
message, say, at global time $t$, while transmitter $u$'s local time is
$\tau_{tu}=t+t_u$.  The receiver $v$ receives this message at a 
later time, when its local clock reads $\tau_{rv}=t+t_v+\delta_{u,v}$,
where $\delta_{u,v}$ is a random delay arising from the processing of
the message at both $u$ and $v$.  Some time later, say at $t'$ (global
time), node $v$ sends a message back to $u$, when its local time is
$\tau'_{tv}=t'+t_v$. This message includes the values $\tau_{rv}$ and
$\tau'_{tv}$ in the message body. Receiver $u$ receives this
message at local time $\tau{'}_{ru} = t'+t_u+\delta_{uv}$, where the
delay $\delta_{vu}$ has the same mean as the delay $\delta_{uv}$.
Node $u$ can now estimate the clock offsets as $\zeta_{u,v} =
\frac{1}{2}\left[(\tau'_{ru}-\tau'_{tv})-(\tau_{rv}-\tau_{tu}) \right]
= t_u-t_v + (\delta_{vu}-\delta_{uv})/2$.  The error
$\varepsilon_{u,v}\eqdef (\delta_{vu}-\delta_{uv})/2$ has zero mean as
long as the (unidirectional) delays have the same expected value. The
measured clock offset between $u$ and $v$ is now
$\zeta_{u,v}=t_u-t_v+\varepsilon_{u,v}$, of the
form~\eqref{eq:measurement}. Similarly, the measurement of clock
offsets between nodes $v$ and $w$ is $\zeta_{v,w}=t_v - t_w +
\varepsilon_{v,w}$.}\label{fig:time-sync}
\end{figure*}
}
{


Several applications in sensor and actuator networks lead to
estimation problems where a number of variables are to be estimated
from noisy measurements of the difference between certain pairs of
them.  Consider the problem of localization, where a sensor does not
know its position in a global coordinate system, but can measure its
position relative to a set of nearby nodes. These measurements can be
obtained, for example, from range and angle data but are typically subjected to  large noise (see~Figure~\ref{fig:xy_from_rtheta}).  In particular, two nearby
sensors $u$ and $v$ located in a plane at positions $x_u$ and $x_v$,
respectively, have access to the measurement
\begin{align}\label{eq:measurement}
  \zeta_{u,v} = x_u - x_v + \epsilon_{u,v},
\end{align}
where $\epsilon_{u,v}$ denotes measurement error. The problem of
interest is to use the $\zeta_{u,v}$'s to estimate the positions of
all the nodes in a common coordinate system whose origin is fixed
arbitrarily at one of the nodes. 

\medskip

Similar estimation problems arise in time synchronization~\cite{Karp_03TimeSync,Kumar-timesync-II:06,Kumar-timesync-I:06} and motion consensus in sensor-actuator networks~\cite{BarooahHespanhaDec06a};
see~\cite{PB_JPH_CSM:07,BarooahHespanhaDec06a} for an overview of
these applications. Motivated by these applications, we study the
problem of estimating vector valued variables from noisy measurements
of the difference between them.  In particular,
denoting the variables of interest by $\{x_i:i\in\mathbf{V}\}$ where
$\mathbf{V}:=\{1,2,\dots\}$, we consider problems for which noisy
relative measurements of the form~\eqref{eq:measurement} are
available. The ordered pairs of indices $(u,v)$ for which we have relative
measurements form a set $\mathbf{E}$ that is a (typically strict)
subset of the set $\V\times\V$ of all pairs of indices.  Just with
relative measurements, the $x_u$'s can be determined only up to an
additive constant.  To avoid this ambiguity, we assume that a
particular variable (say $x_o$) is used as the \emph{reference}, which is therefore assumed known. The problem of interest is to estimate the remaining node variables from all the available measurements.

\medskip

The measurement equations \eqref{eq:measurement} can be naturally
associated with a directed graph $\G=(\V,\E)$ with an edge from node
$u$ to $v$ if the measurement $\zeta_{u,v}$ is available.  The graph
$\G$ is called the \emph{measurement graph}, and each vector $x_u$,
$u\in\V$ is called the \emph{$u$-th node variable}. The measurement
noise $\epsilon_{e}, e\in\E$ is assumed zero mean and spatially
uncorrelated, i.e., $\Exp[\epsilon_e]=0 \; \forall e \in\E$ and
$\Exp[\epsilon_e \epsilon_{\bar e}^T] =0$ if $e \neq \bar{e}$.

\ifthenelse{\equal{\PaperORReport}{Paper}}{
\begin{figure}[b]
  \begin{center}
    \psfrag{u}{$u$}    \psfrag{v}{$v$}    \psfrag{w}{$w$}
    \psfrag{r}{$r_{u,v}$}
    \psfrag{th}{$\theta_{u,v}$}
    \psfrag{r2}{$r_{v,w}$}
    \psfrag{th2}{$\theta_{v,w}$}
    \includegraphics[scale=0.5]{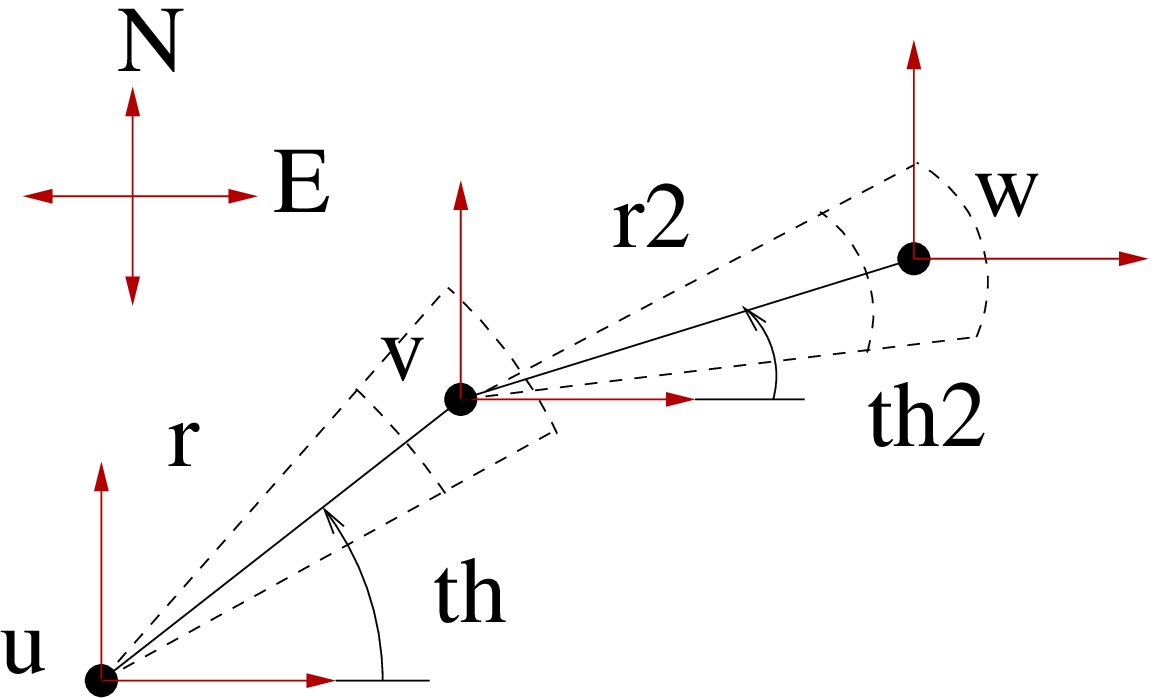}
    \caption{\label{fig:xy_from_rtheta}Relative position measurement
      in a Cartesian reference frame using range and angle
      measurements. A local compass at each sensor is needed to
      measure bearing with respect to a common North.  Noisy
      measurements of the range $r_{u,v}$ and angle $\theta_{u,v}$ between a pair of sensors $u$ and $v$, which are denoted by $\hat{r}_{u,v}$ and $\hat{\theta}_{u,v}$, are converted to noisy
      measurements of relative position in the $x-y$ plane as
      $\zeta_{u,v}=\displaystyle{\frac{1}{\bar{c}}[\hat{r}_{u,v}\cos \hat{\theta}_{u,v}, \,
      \hat{r}_{u,v}\sin \hat{\theta}_{u,v}]^T}$, with $\bar{c} = \Exp [\cos (\delta \theta)]$, where $\delta_\theta = \theta - \hat{\theta}$ is the random error in the angle measurement (see~\cite{PB_JH_scaling_ArXiV:08} for more details). The same procedure is performed      for every pair of sensors that can measure their relative range
      and angle. The task then is to estimate the positions of all the
      nodes with respect to an arbitrary node in the network from the
      relative position measurements.}
  \end{center}
\end{figure}
}
{
\begin{figure}[t]
  \begin{center}
    \psfrag{u}{$u$}    \psfrag{v}{$v$}    \psfrag{w}{$w$}
    \psfrag{r}{$r_{u,v}$}
    \psfrag{th}{$\theta_{u,v}$}
    \psfrag{r2}{$r_{v,w}$}
    \psfrag{th2}{$\theta_{v,w}$}
    \includegraphics[scale=0.5]{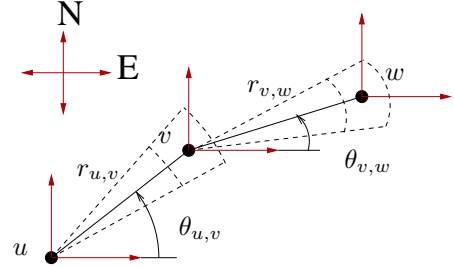}
    \caption{\label{fig:xy_from_rtheta}Relative position measurement
      in a Cartesian reference frame using range and angle
      measurements. A local compass at each sensor is needed to
      measure bearing with respect to a common North.  Noisy
      measurements of the range $r_{u,v}$ and angle $\theta_{u,v}$ between a pair of sensors $u$ and $v$, which are denoted by $\hat{r}_{u,v}$ and $\hat{\theta}_{u,v}$, are converted to noisy
      measurements of relative position in the $x-y$ plane as
      $\zeta_{u,v}=\displaystyle{\frac{1}{\bar{c}}[\hat{r}_{u,v}\cos \hat{\theta}_{u,v}, \,
      \hat{r}_{u,v}\sin \hat{\theta}_{u,v}]^T}$, with $\bar{c} = \Exp [\cos (\delta \theta)]$, where $\delta_\theta = \theta - \hat{\theta}$ is the random error in the angle measurement. The division by $\bar{c}$ is needed to ensure that the noise in the measurement $\zeta_{u,v}$ is zero mean (see Appendix~\ref{sec:unbiased-measurement}). The same procedure is performed      for every pair of sensors that can measure their relative range
      and angle. The task then is to estimate the positions of all the
      nodes with respect to an arbitrary node in the network from the
      relative position measurements.}
  \end{center}
\end{figure}

}

}

\medskip

In this paper we investigate how the structure of the graph $\G$
affects the quality of the optimal linear unbiased estimate $\hat x_u$ of $x_u$,
measured in terms of the covariance of the estimation error
$\Sigma_{u,o} \eqdef \Exp[(x_u - \hat{x}_u)(x_u - \hat{x}_u)^T]$.
The \emph{optimal linear unbiased  estimate} refers to the one obtained with the classical best linear unbiased estimator (BLUE), which achieves the minimum
variance among all linear unbiased estimators~\cite{tutorial_Rhodes_TAC:71}. We examine the growth of the BLUE error variance of a node $u$ as a function of its distance to the reference node.

\medskip

We are interested in the growth of error with distance in large graphs, for which infinite graphs (with a countably infinite number of nodes and edges) serve as proxies. This paper is focused on infinite graphs because the absence of boundary conditions in infinite graphs allows for more complete and simpler results. Using infinite graphs as proxies for large finite graphs is theoretically justified by the fact that the BLUE error variance of a node variable $x_u$ in a large but finite subgraph of an infinite graph is arbitrarily close to the BLUE estimation error in the infinite graph, as long as the finite graph is sufficiently large. This convergence result was established in~\cite{PB_JH_TSP:08}. 

\medskip

When the measurement graph is a tree, there is a single path between
the $u^\text{th}$ node and the reference node and one can show that the
covariance matrix of the estimation error is the sum of the covariance
matrices associated with this path.  Thus, for trees, the variance of
the BLUE estimation error of $x_u$ grows linearly with the distance
from node $u$ to the reference node. It turns out that for graphs
``denser'' than trees, with multiple paths between pairs of nodes, the
variance of the optimal linear unbiased estimation error can grow slower than linearly with distance. 

\medskip

In this paper, we introduce a novel notion of denseness for graphs that is needed to characterize how the estimation error grows with distance. In
classical graph-theoretic terminology, a graph with $n$ vertices is
called dense if its average node degree is of order $n$, and is called
sparse if its average node degree is a constant independent of
$n$~\cite{Diestel_GT:05}. We recall that the degree of a node is the number of edges
incident on it (an edge $(u,v)$ is said to be incident on the nodes
$u$ and $v$). Other notions of denseness include geo-denseness introduced by~\cite{CA_GE_TCS:07}, which requires uniform node density (nodes per unit area) but does not consider the role of edges. Accuracy of localization from distance-only measurements have been  extensively studied in the sensor networks literature, typically by evaluating the Cram{\'e}r-Rao lower bound (see~\cite{Pat03_MLECR,DN_BN_MOBIHOC:04,AS_WG_RM_MS_TMC:04,CC_AS_EURASIP:06,NA_KP_EURASIP:08} and references therein). In many of these studies, graph density (as measured by node degree or node density) is recognized to affect estimation accuracy~\cite{DN_BN_MOBIHOC:04,AS_WG_RM_MS_TMC:04,NA_KP_EURASIP:08}. However, we will see through examples in Remark~\ref{rem:counterexamples} that for the estimation problem considered in this paper, such notions of denseness are not sufficient to characterize how the estimation error grows with distance.

\medskip

\begin{figure}
  \begin{center}
    \hfill
    \subfigure[\normalsize{1D lattice $\Z_1$}]{
      \includegraphics[width=.3\columnwidth]{./epsfiles/1dlatt.eps}}
    \hfill
    \subfigure[\normalsize{2D lattice $\Z_2$}]{
      \includegraphics[width=.3\columnwidth]{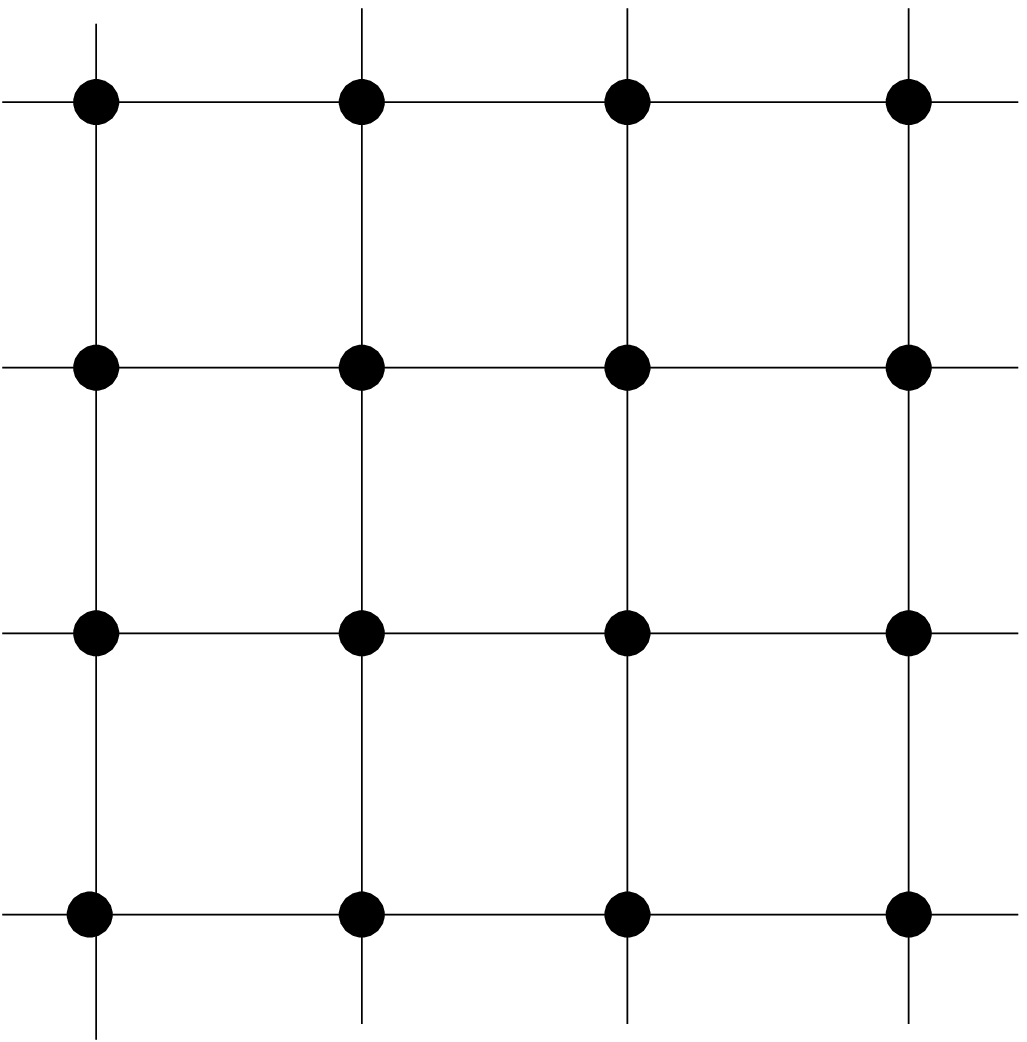}}
    \hfill
    \subfigure[\normalsize{3D lattice $\Z_3$}]{
      \includegraphics[width=.3\columnwidth]{./epsfiles/3dlatt.eps}}
    \hspace*{\fill}
  \end{center}
\caption{\label{fig:lattices}Lattices.}
\end{figure}

A key contribution of this paper is the development of suitable notions of graph denseness and sparseness that are useful in
determining BLUE error scaling laws. These notions exploit the
relationship between the measurement graph and a lattice. We recall that the $d$-dimensional square lattice $\Z_d$ is defined as a graph with a node in every point in $\R^d$ with integer coordinates and an edge between every pair of
nodes at an Euclidean distance of $1$ (see Figure~\ref{fig:lattices} for examples). The error scaling laws for a lattice measurement graph can be determined analytically by exploiting  symmetry. It turns out that when the graph is not a lattice, it can still be compared to a lattice. Intuitively, if after some
bounded perturbation in its node and edge set, the graph looks approximately
like a $d$-dimensional lattice, then the graph inherits the denseness properties of the lattice. In that case the error covariance for the lattice can still be used to bound the error covariance in the original graph.

\medskip

Our classification of dense and sparse graphs in
$\R^d$, $d \in \{1,2,3\}$, characterizes BLUE error scaling laws. For
dense graphs, they provide upper bounds on the growth rate of the error, while for sparse graphs, they provide lower bounds. The precise growth rates depend on which dimension the graph
is dense or sparse in. When a graph is dense in $1$D, $2$D, or $3$D,
respectively, the error covariance of a node is upper bounded by a
linear, logarithmic, or bounded function, respectively, of its distance from the
reference.  On the other hand, when a graph is sparse in $1$D, $2$D,
or $3$D, the error covariance of a node is lower bounded
by a linear, logarithmic, and bounded function,  respectively, of its distance from
the reference. Our sparse graphs are also known as ``graphs that can be drawn in a civilized manner'' according to the terminology introduced by~\citet{DoyleSnell} in connection with random walks.

\smallskip

The BLUE error scaling laws derived in this paper provide an
algorithm-independent limit to the estimation accuracy achievable in
large networks, since no linear unbiased estimation algorithm can achieve higher
accuracy than the BLUE estimator. For example, when a graph is
sparse in $1$D, the BLUE estimation error covariance grows at least
linearly with the distance from the reference. Therefore the
estimation accuracy will be necessarily poor in large 1D sparse graphs. On the
other hand, when a graph is dense in $3$D, the BLUE estimation error of every node variable remains below a constant, even for nodes that are arbitrarily far away from the reference. So accurate estimation is possible in very large 3D dense graphs.

\smallskip

The results in this paper are useful for the design and deployment
 of ad-hoc and sensor networks. Since we now know what structural
properties are beneficial for accurate estimation, we can strive to
achieve those structures when deploying a network. Specifically, we should try to achieve a dense-in-$\R^d$ structure, with $d$ as large as possible, for high accuracy estimation. Since the scaling laws are true for the optimal linear unbiased estimator, they can also help designers determine if design requirements are achievable. For example, if the requirement is that the estimation accuracy should not decrease with size, no matter how large a network is, the network must be dense in $\R^d$, $d\geq 3$ for such a requirement to be satisfied.

\smallskip

Our results also expose certain misconceptions that exist in
the sensor network literature about the relationship between graph
structure and estimation error. In Section~\ref{sec:scalnglaws},
we provide examples that expose the inadequacy of the usual measures
of graph denseness, such as node degree, in determining scaling laws
of the estimation error.

\smallskip

In practice, more than one reference node (commonly referred to as anchors) may be used. We only consider the case of a single reference node since scaling laws with a single reference provide information on how many reference nodes may be needed. For example, since the estimation error in a 3D dense graph is bounded by a constant, one reference node may be enough for such a graph.

\smallskip

While we do not discuss the computation of the optimal linear unbiased estimates in this paper, we have developed distributed algorithms to compute these estimates with arbitrary precision (see~\cite{PB_JPH_CSM:07} and references therein). These algorithms are distributed in the sense that every node computes its own estimate and the information needed to carry out this computation is obtained by communication with its neighbors.

\smallskip

A preliminary version of some of the results in this paper was presented in~\cite{PB_JPH:05}. However,~\cite{PB_JPH:05} used stricter assumptions to establish the upper bounds on error growth rates. Moreover, only sufficient conditions were obtained in~\cite{PB_JPH:05} for some of the error scaling laws to hold; whereas here we derive necessary and sufficient conditions.

\smallskip

\emph{Organization:} The rest of the paper is organized as
follows. Section~\ref{sec:mainresults} describes the problem and summarizes the main results of the paper. Section~\ref{sec:densesparse-more} describes key properties of dense and sparse graphs. Section~\ref{sec:analogy} briefly describes the analogy between BLUE and
generalized electrical networks from~\cite{PB_JH_TSP:08} that is needed to prove the main results. Section~\ref{sec:mainproof} contains
the proof of the main result of the paper. Section~\ref{sec:checking-denseness} deals with the question of how to check if a graph possesses the denseness/sparseness properties. The paper ends with a a few final conclusions and directions for future research in Section~\ref{sec:summary}.

\section{Problem Statement and Main Results}\label{sec:mainresults}
Recall that we are interested in estimating vector-valued variables
$x_u \in \R^k$, $u\in\V\eqdef \{1,2,\dots\}$, from noisy relative
measurements of the form:
\begin{align}\label{eq:measurement2}
\zeta_{u,v}= x_u - x_v+ \epsilon_{u,v}, \quad (u,v)\in\E
\end{align}
where $\epsilon_{u,v}$ denotes a zero-mean measurement noise and $\E$
is the set of ordered pairs $(u,v)$ for which relative measurements
are available. The node set $\V$ is either finite, or infinite but
countable. We assume that the value of a particular \emph{reference
variable} $x_o$ is known and without loss of generality we take $x_o =
0$. The node set $\V$ and the edge set $\E$ together define a directed
measurement graph $\G=(\V,\E)$.

\medskip

The accuracy of a node variable's estimate, measured in terms of the
covariance of the estimation error, depends on the graph $\G$ as well
as the measurement errors. The covariance matrix of the
error $\epsilon_{u,v}$ in the measurement $\zeta_{u,v}$ is denoted by
$P_{u,v}$, i.e., $P_{u,v} \eqdef \Exp[\epsilon_{u,v} \epsilon_{u,v}^T]
$. We assume that the measurement errors on different edges are uncorrelated, i.e.,
for every pair of distinct edges $e,\bar e \in \E$, $\Exp[\epsilon_e
\epsilon_{\bar e}^T]=0$.  The estimation problem is now formulated
in terms of a \emph{network} $(\G,P)$ where $P:\E \to \S^{k+}$ is a
function that assigns to each edge $(u,v)\in\E$ the  covariance
matrix $P_{u,v}$ of the measurement error associated with the edge
$(u,v)$ in the measurement graph $\G$. The symbol $\S^{k+}$ denotes
the set of $k \times k$ symmetric positive-definite matrices.

\medskip

As discussed in Section~\ref{sec:intro}, our results are stated for infinite networks. The following conditions are needed to make sure that the estimation problem is well posed and that the estimates satisfy appropriate convergence properties to be discussed shortly:

\medskip

\begin{assumption}[measurement network]\label{as:bounded}
  The measurement network $(\G,P)$ satisfies the following properties:
  \begin{enumerate}
  \item The graph $\G$ is weakly connected, i.e., it is possible to go from every node to every other node traversing the graph edges without regard to edge direction.
  \item The graph $\G$ has a finite maximum node
  degree\footnote{The degree of a node is the number of edges that are incident on the node. An edge $(u,v)$ is said to be incident on the nodes $u$ and $v$.}.
  \item The edge-covariance function $P$ is uniformly
  bounded, i.e., there exists constant symmetric positive matrices 
  $P_{\min},P_{\max}$ such that $ P_{\min} \leq P_e
  \leq P_{\max}$, $\forall e \in \E$. \frqed
 \end{enumerate}
\end{assumption}
In the above, for two matrices $A,B\in\R^{k\times k}$, $A>B$ ($A \ge B$)
means $A-B$ is positive definite (semidefinite). We write $A<B$ ($A \le B$) if $-A > -B$ ($-A \ge -B$). 

\medskip

We also assume throughout the paper that measurement graphs do not have parallel edges. A number of edges are said to be parallel if all of them are incident on the same pair of nodes. The condition of not having parallel edges is not
restrictive since parallel measurements can be combined into a
single measurement with an appropriate covariance, while preserving the BLUE
error covariances (see Remark~\ref{rem:parallel}). 

\medskip

Given a finite measurement network $(\G_\mrm{finite},P)$, where 
$\G_\mrm{finite}$ contains the nodes $u$ and $o$, it is
straightforward to compute the BLUE estimate
$\hat{x}_u(\G_\mrm{finite})$ of the unknown node variable $x_u$ in the
network $(\G_\mrm{finite},P)$, as described in~\cite{PB_JH_TSP:08}, and the covariance matrix of the estimation error $\Sigma_{u,o}(\G_\mrm{finite}) \eqdef
\Exp[(x_u - \hat{x}_u)(x_u - \hat{x}_u)^T]$ exists as long as 
$\G_\mrm{finite}$ is weakly
connected~\cite{PB_JH_TSP:08}. Due to the
optimality of the BLU estimator, $\Sigma_{u,o}(\G_\mrm{finite})$ is
the minimum possible estimation error covariance that is achievable by any linear unbiased estimator using all the measurements in the graph $\G_\mrm{finite}$.

\medskip

When the measurement graph is infinite, the BLUE error covariance
$\Sigma_{u,o}$ for a node variable $x_u$ is defined as
\begin{align}\label{eq:Sigma-infinite-defn}
\Sigma_{u,o} & = \inf_{\G_\mrm{finite}} \Sigma_{u,o}(\G_\mrm{finite}),
\end{align}
where the infimum is taken over all finite subgraphs $\G_\mrm{finite}$
of $\G$ that contain the nodes $u$ and $o$.  We define a matrix $M$ to be the \emph{infimum} of the matrix set $S \subset \S^{k+}$, and denote it by 
\begin{align}\label{eq:matrix-inf-defn}
M = \inf S,
\end{align}
if $M\leq A$ for every matrix $A \in S$, and for every positive
real $\epsilon$, there exists a matrix $B \in S$ such that $M+\epsilon I_k >
B$. Under Assumption~\ref{as:bounded}, it was shown in~\cite{PB_JH_TSP:08} that the infimum in~\eqref{eq:Sigma-infinite-defn} always exists. In this case,~\eqref{eq:Sigma-infinite-defn} means that the BLUE covariance $\Sigma_{u,o}$ is the the lowest error covariance that can be achieved by using all the available measurements.

\medskip

In the sequel, we determine how the BLUE covariance $\Sigma_{u,o}$ grows as a function of the distance
of node $u$ to the reference $o$, and how this scaling law depends on the structure of the measurement graph $\G$. To this effect we start by providing a classification of graphs that is needed to characterize the error scaling laws.


\subsection{Graph Denseness and Sparseness}\label{sec:drawing}
We start by introducing graph drawings, which will later allow us to define dense and sparse graphs.

\subsubsection{Graph Drawings}
The drawing of a graph $\G=(\V,\E)$ in a $d$-dimensional Euclidean
space is obtained by mapping the nodes into points in $\R^d$ by a
\emph{drawing function} $f:\V \to \R^d$. A drawing is also called a
representation~\cite{GodsilRoyle_2001} or an embedding~\cite{Diestel_GT:05}.  For a particular drawing $f$, given two nodes $u,v\in
\V$ the \emph{Euclidean distance between $u$ and $v$ induced by the
  drawing $f:\V \to \mathbb{R}^d$} is defined by
\begin{align*}
  d_f(u,v) := \| f(v)-f(u) \|,
\end{align*}
where $\|\cdot\|$ denoted the usual Euclidean norm in $\R^d$. It is
important to emphasize that the definition of drawing allows
edges to intersect and therefore every graph has a drawing in
every Euclidean space. In fact, every graph has an infinite number of
drawings in every Euclidean space. However, a particular drawing 
is useful only if it clarifies the relationship between the
graph and the Euclidean space in which it is drawn. In what follows, given two nodes $u$ and $v$, $d_\G(u,v)$ denotes the \emph{graphical distance} between $u$ and $v$, i.e., the number of edges in the shortest path between $u$ and $v$. The graphical distance $d_\G$ is evaluated without regards to edge directions, which are immaterial in determining BLUE error covariances (see Remark~\ref{rem:Reff-directions}).


\medskip

For a particular drawing $f$ and induced Euclidean distance $d_f$ of a
 graph $\G=(\V,\E)$, four parameters are needed to characterize graph
 denseness and sparseness. The \emph{minimum node distance}, denoted
 by $s$, is defined as the minimum Euclidean distance between the
 drawing of two nodes
\begin{align*}
  s:= \inf_{\substack{u,v\in\V\\v\neq u}} d_f(u,v).
\end{align*}
The \emph{maximum connected range}, denoted by $r$, is defined as the
Euclidean length of the drawing of the longest edge
\begin{align*}
  r:= \sup_{(u,v)\in\E} d_f(u,v).
\end{align*}
The \emph{maximum uncovered diameter}, denoted by $\gamma$, is defined
as the diameter of the largest open ball that can be placed in $\R^d$
such that it does not enclose the drawing of any node
\begin{align*}
  \gamma:= \sup \Big\{\delta : \exists \ball_\delta \text{ s.t. }
  f(u)\notin \ball_\delta, \forall u\in\V\Big\},
\end{align*}
where the existential quantification spans over the balls $\ball_\delta$
in $\R^d$ with diameter $\delta$ and centered at arbitrary points.
Finally, the \emph{asymptotic distance ratio}, denoted by $\rho$, is
defined as
\begin{align*}
  \rho := \lim_{n\to \infty} \inf \Big\{\frac{d_f(u,v)}{d_\G(u,v)}: u,v\in\V
  \text{ and } d_\G(u,v)\geq n\Big\}.
\end{align*}
Essentially $\rho$ provides a lower bound for the
ratio between the Euclidean and the graphical distance for nodes that
are far apart. The asymptotic distance ratio can be thought of as an
inverse of the \emph{stretch} for geometric graphs, which is a
well-studied concept for finite graphs~\cite{GN_MS_SIAM:00}.

\begin{figure}[ht]
\psfrag{r}{$r$}
\psfrag{s=1}{$s=1$}
\psfrag{g}{$\gamma$}
\psfrag{u1}{$u^*$}\psfrag{v1}{$v^*$}
\psfrag{u2}{$p^*$}\psfrag{v2}{$q^*$}
\begin{center}
\includegraphics[width=2in]{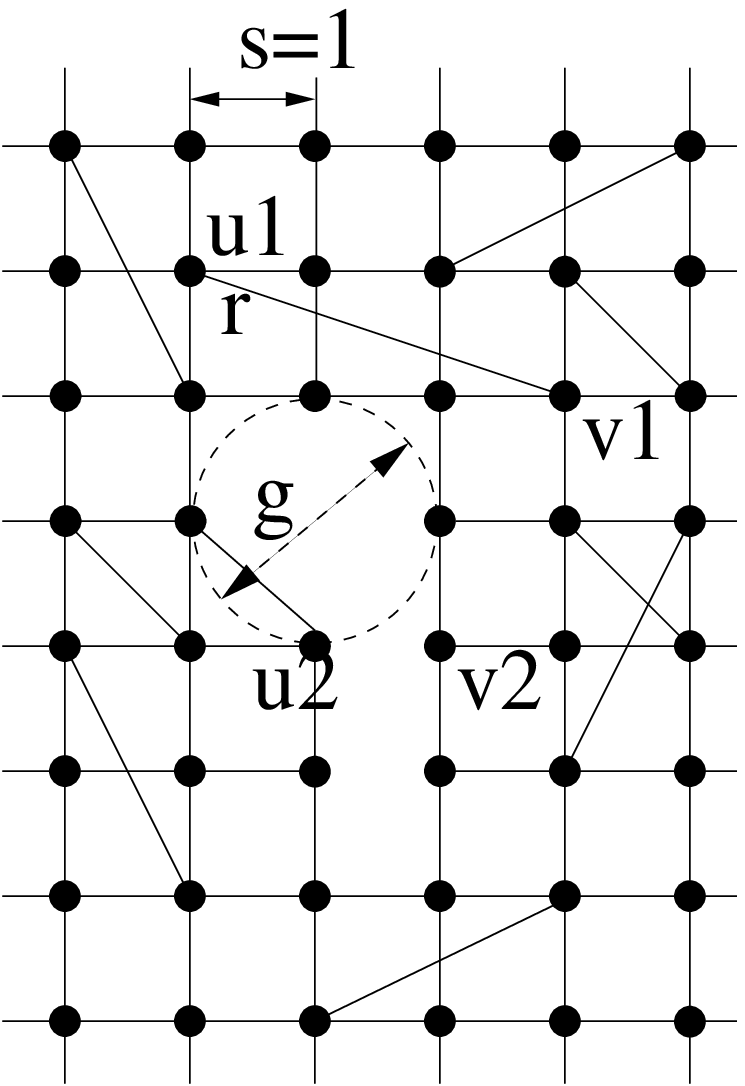}
\caption{\label{fig:drawing_params_def}A drawing of a graph in 2D Euclidean space, and the corresponding denseness and sparseness
  parameters. Since the minimal distance between any two nodes is $1$, the minimum node distance is $s=1$. Since the longest edge is between
  $u^*$ and $v^*$, the maximum connected
  range is $r=\sqrt{10}$. The diameter of the largest ball that can
  fit inside the drawing without enclosing any node is $2$, so the
   maximum uncovered diameter is thus $\gamma = 2$. The minimal ratio
  between the Euclidean and graphical distance of a pair of nodes is
  achieved by the pair $p^*,q^*$, hence the asymptotic distance
  ratio is $\rho = d_f(p^*,q^*)/d_\G(p^*,q^*) = 1/5$.}
\end{center}
\end{figure}

\medskip

The two parameters $\rho$ and $r$ defined above are especially useful to compare graphical and Euclidean distances, as stated in the following result.

\begin{lemma}[Euclidean vs.~graphical distances]\label{lem:rho}
  The following two statements are equivalent:
  \begin{enumerate}
  \item\label{en:rho} The asymptotic distance ratio $\rho$ is strictly positive.
  \item\label{en:alpha-beta} There exist constants $\alpha > 0 ,\beta >0$ for
    which
    \begin{align}\label{eq:rho-a-b}
      d_\G(u,v) \leq \alpha d_f(u,v)+\beta,\quad \forall u,v \in \V. 
    \end{align}
  \end{enumerate}
  Similarly, the following statements are equivalent:
  \begin{enumerate}
  \item\label{en:r} The maximum connected range $r$ is finite.
  \item\label{en:alpha-beta-r} There exist constants $\alpha>0$, $\beta \geq 0 $ for
    which
    \begin{align*}
      d_f(u,v) \leq \alpha d_\G(u,v)+\beta,\quad \forall u,v \in \V. \tag*{\frqed}
    \end{align*}
  \end{enumerate}
\end{lemma}

\ifthenelse{\equal{\PaperORReport}{Paper}}{
The proof of this lemma is provided in the Appendix.
}{
The proof of this lemma is provided in Appendix~\ref{sec:App-proofs-1}.
}

\subsubsection{Dense and Sparse Graphs}\label{sec:dense-sparse-defn}
We call the drawing of a graph with finite maximum uncovered diameter ($\gamma<\infty$) and positive asymptotic distance ratio ($\rho>0$) a \emph{dense drawing}.  We say that a graph \emph{$\G$ is
  dense in $\R^d$} if there exists a dense drawing of the graph in
$\R^d$.  Graph drawings for which the minimum node distance is positive ($s>0$)
and the maximum connected range is finite ($r<\infty$) are called
\emph{civilized drawings}~\cite{DoyleSnell}. A graph $\G$ is said to
be \emph{sparse in $\R^d$} if there exists a civilized drawing in
$\R^d$.


It follows from these definitions and Lemma~\ref{lem:rho} that if a graph is dense in $\R^d$, then it has enough nodes and edges so that it is possible to draw it in $\R^d$ in such a way that its nodes cover $\R^d$ without leaving large holes (finite $\gamma$), and yet a small Euclidean distance between two nodes in the drawing guarantees a small graphical
distance between them (positive $\rho$, which implies~\eqref{eq:rho-a-b}). On the other hand, if a graph that is sparse in $\R^d$, then one can draw it in $\R^d$ so as to keep a certain minimum separation between nodes (positive $s$) without making the edges arbitrarily long (finite $r$). It also follows from the definitions that a graph must be infinite to be dense in any dimension, and a finite graph is sparse in every dimension.

\smallskip

A graph can be both dense and sparse in the same dimension.  For
example, the $d$-dimensional lattice is both
sparse and dense in $\R^d$. However, there is no civilized drawing of
the $d$-dimensional lattice in $\R^{d'}$ for any $d'<d$. Moreover, there is no dense drawing of the $d$-dimensional
lattice in $\R^{\bar d}$ for every $\bar d > d$.  This means, for
example, that the 3D lattice in not sparse in 2D and is not dense in
4D. In general, a graph being dense in a particular dimension puts a
restriction on which dimensions it can be sparse in. The next result,
proved in Section~\ref{sec:checking-denseness}, states this precisely.

\medskip

\begin{lemma}\label{lem:noconfusion}
A graph that is dense in $\R^d$ for some $d\geq 2$, cannot be sparse
in $\R^{d'}$ for every $d'<d$. \frqed
\end{lemma}

\begin{table*}[t!]
  \begin{center}
\caption{\label{tab:growth} Covariance $\Sigma_{u,o}$ of
  $x_u$'s BLUE estimation error for graphs that are dense or sparse in
  $\R^d$. In the table, $d_{f_d}(u,o)$ denotes the Euclidean distance
  between node $u$ and the reference node $o$ induced by a drawing $f_d:\V
  \to \R^d$ that establishes the graph's denseness in the
  Euclidean space $\R^d$, and $d_{f_d'}(u,o)$ denotes the Euclidean distance induced by a drawing $f_{d}': \V \to \R^{d}$ that establishes the graph's sparseness.}
    \begin{tabular}{c|l|l}
      \parbox{1.2 in }{\begin{center}Euclidean space \\ and graph example \end{center}} &
      \parbox{2.3in}{\begin{center}Covariance matrix $\Sigma_{u,o}$ of the estimation error of
                     $x_u$ in a \emph{sparse graph} with a sparse drawing $f_{d}'$ \end{center}} &
      \parbox{2.3in}{\begin{center}Covariance matrix $\Sigma_{u,o}$ of the estimation error of
                      $x_u$ in a \emph{dense graph} with a dense drawing $f_{d}$\end{center}}
      \\[3ex]\hline&&\\[-2ex]
      \includegraphics[trim=0 -.5in 0in -.5in,width=.35in]{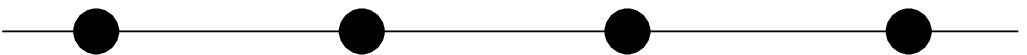}\quad
      $\R$ & \normalsize $ \Sigma_{u,o}(\G) = \Omega\Big( d_{f_1'}(u,o)\Big)$
      & \normalsize $\Sigma_{u,o}(\G) = \scr{O}\Big(d_{f_1}(u,o)\Big)$ \\
      \includegraphics[width=.35in]{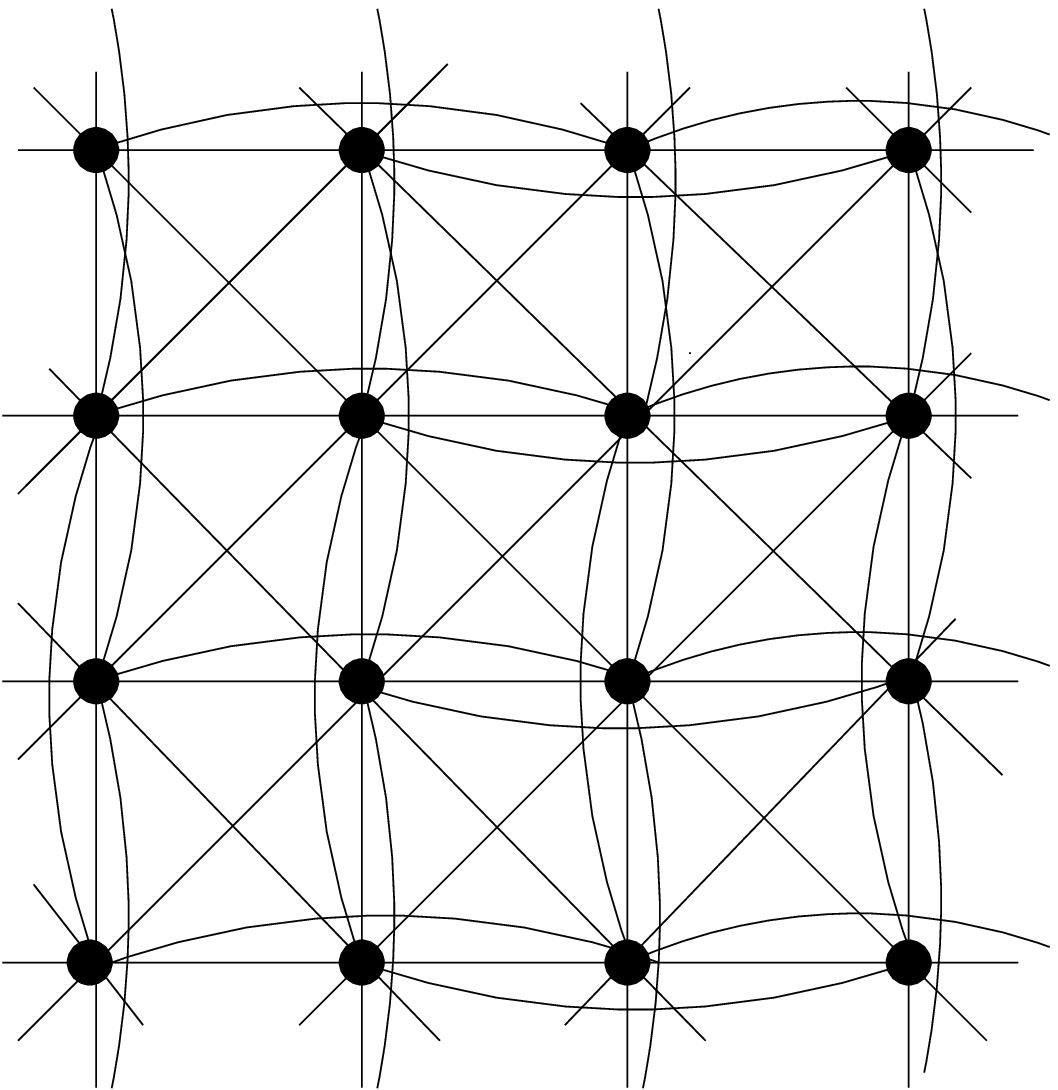}\quad
      $\R^2$ & \normalsize $\Sigma_{u,o}(\G) = \Omega\Big(\log d_{f_2'}(u,o)\Big)$
      & \normalsize $\Sigma_{u,o}(\G) = \scr{O}\Big(\log d_{f_2}(u,o)\Big)$ \\
      \includegraphics[width=.35in]{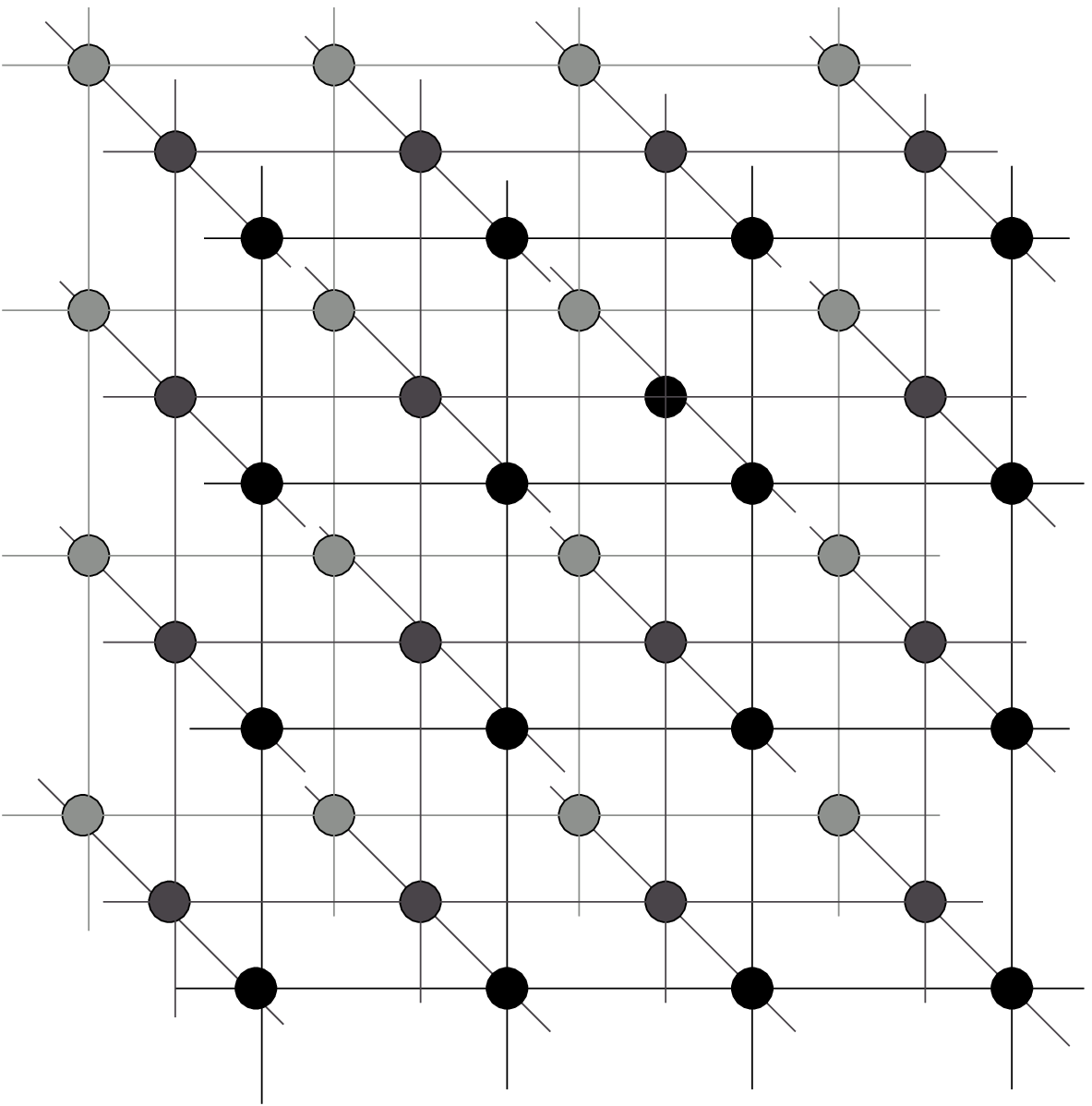}\quad
      $\R^3$ & \normalsize $\Sigma_{u,o}(\G)=\Omega\Big(1\Big)$
      & \normalsize $ \Sigma_{u,o}(\G) = \scr{O}\Big(1\Big)$
    \end{tabular}\\ \vspace{2 ex}
  \end{center}
\end{table*}

\smallskip

\begin{remark}[historical note] In the terminology of~\citet{DoyleSnell}, sparse graphs (as defined here) are said to be graphs ``that can be drawn in a civilized manner''. In this paper we refer to such graphs as sparse graphs since they are the antitheses of dense graphs. \frqed
\end{remark}

\subsection{Error Scaling Laws}\label{sec:scalnglaws}
The concepts of dense and sparse graphs allow one to characterize
precisely how the BLUE error covariance $\Sigma_{u,o}$ grows with the
distance from the node $u$ to the reference $o$.  The next theorem, which establishes the BLUE error scaling laws for dense and sparse graphs, is the main result of the paper. The proof of the theorem is provided in
Section~\ref{sec:mainproof}.

\medskip

Before we present the theorem, we need to introduce some notation. The asymptotic notations $\Omega(\cdot)$ and $\scr{O}(\cdot)$ are used for matrix valued
functions in the following way. For a matrix-valued function $g :
\R \to \R^{k \times k}$ and a scalar-valued function $p:\R \to \R$,
the notation $g(x) = \scr{O} (p(x))$ means that there exists a
positive constant $x_o$ and a constant matrix $A\in\S^{k + }$ such
that $g(x) \leq A p(x)$ for all $x > x_o$.  Similarly, $g(x) =
\Omega(p(x))$ means there exists a positive constant $x_o$ and a
constant matrix $B\in\S^{k+}$ such that $g(x) \geq B p(x)$ for all $x
> x_o$. Recall that $\S^{k+}$ is the set of all $k \times k $
symmetric positive definite matrices.

\medskip

\begin{theorem}[Error Scaling Laws]\label{thm:errorscalinglaws}
  Consider a measurement graph $\G=(\V,\E)$ that satisfies
  Assumption~\ref{thm:errorscalinglaws}, with a reference
  node $o\in\V$. The BLUE error covariance $\Sigma_{u,o}$ for a
  node $u$ obeys the scaling laws shown in
  Table~\ref{tab:growth}. \frqed
\end{theorem}

\medskip

A graph can be both sparse and dense in a particular
dimension, in which case the asymptotic upper and lower bounds are the
same. For a graph that is both sparse and dense in $\R^d$, the error covariance grows with distance in the same rate as it does in the corresponding lattice $\Z_d$. 

\medskip

\begin{remark}[Counterexamples to conventional wisdom]\label{rem:counterexamples}
As noted in Section~\ref{sec:intro},  the average node degree of a graph or the number of nodes and edges per unit area of a deployed network are often used as  measures of graph denseness. However, these measures do not predict error scaling
laws.  The three graphs in Figure~\ref{fig:trex} offer an example of
the inadequacy of node degree as a measure of denseness. This figure shows a
$3$-fuzz of the 1D lattice (see Section~\ref{sec:densesparse-more} for
the formal definition of a fuzz), a triangular
lattice\index{triangular lattice}, and a $3$-dimensional lattice. It can be verified from the definitions in
Section~\ref{sec:dense-sparse-defn} that the $3$-fuzz of the 1D
lattice is both dense and sparse in $\R$, the triangular lattice is
dense and sparse in $\R^2$, and the 3D lattice is dense and sparse in
$\R^3$. Thus, it follows from Theorem~\ref{thm:errorscalinglaws} that
the BLU estimation error scales linearly with distance in the $3$-fuzz
of the $1$D lattice, logarithmically with distance in the
triangular lattice, and is uniformly bounded with respect to distance
in the $3$D lattice, \emph{even though every node in each of these graphs
has the same degree, namely six}. \frqed \draftnote{\textbf{Joao wanted to remove the ``Remark'' heading, but is kept since it is referred to in the introduction. }}

\begin{figure}
  \begin{center}
    \begin{tabular}{c}
    \subfigure[\normalsize{A $3$-fuzz of a 1D lattice}]{
      \includegraphics[clip=true, trim = -1in 0 -1in 0, scale = 0.3]{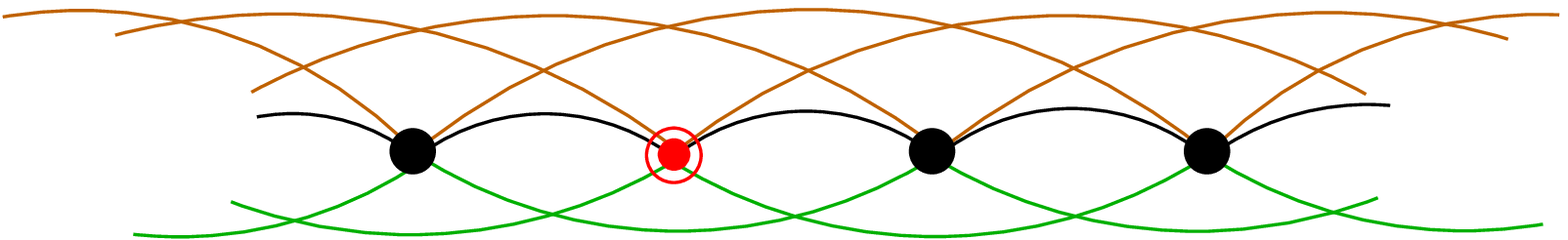} }
    \\
    \subfigure[\normalsize{A triangular lattice}]{
      \includegraphics[clip=true, trim = -1in 0 -1in 0, scale = 0.2]{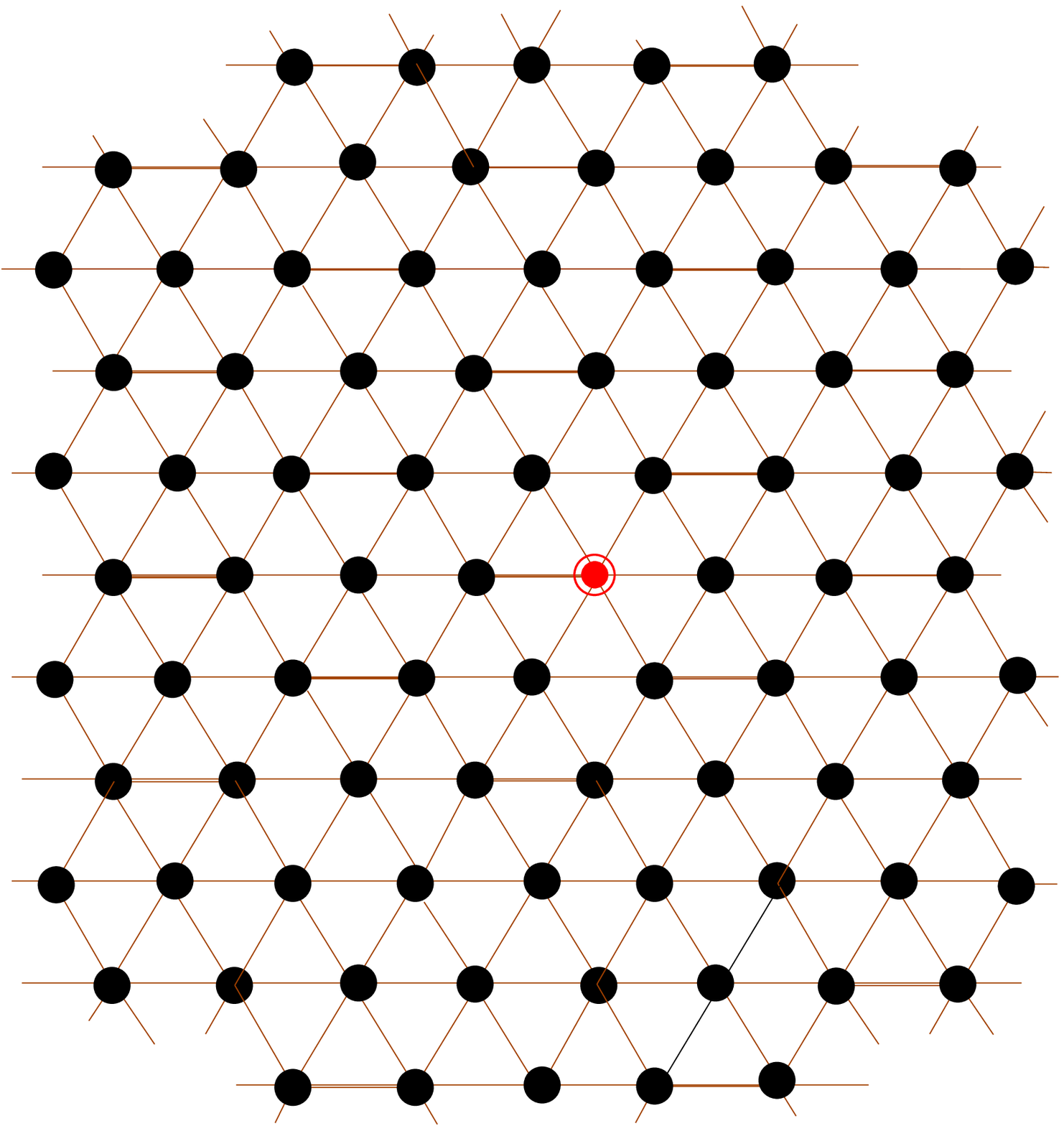}}
    \\
        \subfigure[\normalsize{A 3D lattice}]{
      \includegraphics[clip=true, trim = -1in 0 -1in 0, scale = 0.3]{./epsfiles/3dlatt.eps}}
    \end{tabular}
    \caption{\label{fig:trex}Three measurement graphs that show vastly
      different scaling laws of the estimation error, whereas each has
      the same node degree for every node. Furthermore, they are all
      ``sparse'' according to traditional graph-theoretic terminology
      (see the discussion on graph denseness in
      Section~\ref{sec:intro}). }
  \end{center}
\end{figure}
We note that the notion of geo-denseness introduced in~\cite{CA_GE_TCS:07} is also not useful for characterizing error scaling laws since geo-denseness  considers node density alone without regard to the edges.\end{remark}

\section{Dense and Sparse Graphs}\label{sec:densesparse-more}
This section establishes an embedding relationship between dense and sparse graphs and lattices, which is needed to prove Theorem~\ref{thm:errorscalinglaws}. Roughly speaking, a graph $\G$ can be embedded in another
graph $\bar{\G}$ if $\bar{\G}$ contains all the nodes and edges of
$\G$, and perhaps a few more.  The usefulness of embedding in answering the error scaling
question is that when $\G$ can be embedded in $\bar\G$, the BLUE error
covariances in $\G$ are larger than the corresponding ones in $\bar\G$
(this statement will be made precise in Theorem~\ref{thm:Rayleigh} of
Section~\ref{sec:analogy}). 


\smallskip

 The
\emph{$h$-fuzz of a graph $\G$,} introduced by~\citet{DoyleSnell}, is a
graph with the same set of nodes as $\G$ but with a larger set of
edges. Specifically, given a graph $\G$ and a positive integer $h$, a $h$-fuzz of $\G$, denoted by $\G^{(h)}$, is a graph that has an edge between two
nodes $u$ and $v$ whenever the graphical distance between these nodes in $\G$
is less than or equal to $h$. 

\smallskip

We say that a graph $\G=(\V,\E)$ can be \emph{embedded} in
another graph $\bar{\G}=(\bar{\V}, \bar{\E})$ if $\V \subset \bar{\V}$, and,
whenever there is an edge between two nodes in $\G$, there is an edge
between them in $\bar{\G}$. More precisely, $\G$ can be embedded in
$\bar\G$ if there exists an injective map $\eta:\V \to \bar\V$ such that 
for every $(u,v)\in\E$, either $(\eta(u),\eta(v))\in\bar\E$ or $(\eta(v),\eta(u))\in\bar\E$. In the sequel, we use $\G \subset \bar\G$ to denote that $\G$ can be embedded in $\bar\G$. 

\medskip

\subsection{Relationship with lattices and Euclidean spaces}
The next theorems (Theorem~\ref{thm:embedding-sparse} and~\ref{thm:embedding-dense}) show that  sparse graphs can be embedded in fuzzes of
Lattices, and fuzzes of dense graphs can embed lattices. In these two theorems we use $d_{\Z_d}(\cdot)$ to denote the graphical distance in the lattice $\Z_d$ and $d_f(\cdot)$ to denote the Euclidean distance in $\R^d$ induced by the drawing $f$.

\smallskip

\begin{theorem}[Sparse Embedding]\label{thm:embedding-sparse}
  A graph $\G=(\V,\E)$ 
  is sparse in $\R^d$ if and only if there exists a positive integer
  $h$ such that $\G \subset \Z_d^{(h)}$.  Moreover, if $f:\V \to \R^d$
  is a civilized drawing of $\G$ in $\R^d$, then there exists an
  embedding $\eta:\V \to \V_{\Z_d}$ so that $\forall u,v \in\V$,
\begin{align}\label{eq:civil-embedding-distancerelation}
d_{\Z_d}(\eta(u), \eta(v)) \geq \sqrt{d}\left(\frac{1}{s}d_f(u,v) - 2\right). 
\end{align}
where $s$ is the minimum node distance in the $f$-drawing of $\G$. \frqed
\end{theorem}

\smallskip

In words, the theorem states that $\G$ is sparse in $\R^d$ if and only
if $\G$ can be embedded in an $h$-fuzz of a $d$-dimensional
lattice. The significance of the additional
condition~\eqref{eq:civil-embedding-distancerelation} is that if the
Euclidean distance between a pair of nodes $u$ and $v$ in a civilized
drawing of the graph is large, the graphical distance in the lattice
between the nodes that correspond to $u$ and $v$ must also be large.

\smallskip

The first statement of Theorem~\ref{thm:embedding-sparse} is essentially taken from~\cite{DoyleSnell}, where it was proved that if a graph can be drawn in a civilized manner in $\R^d$, then it can be embedded in a
$h$-fuzz of a $d$-lattice, where $h$ depends only on $s$ and $r$. A
careful examination of the proof in~\cite{DoyleSnell} reveals that it is not only sufficient but also a necessary condition for embedding in lattice
fuzzes. The proof of this theorem is therefore omitted.

\medskip

\begin{theorem}[Dense Embedding]\label{thm:embedding-dense}
 A graph $\G=(\V,\E)$ is dense in $\R^d$ if and only if there exists
 finite, positive integers $h$ and $c$ such that the following
 conditions are satisfied
 \begin{enumerate}[(i)]
 \item $\G^{(h)} \supset \Z_d$,  and, 
 \item if $\eta:\V_{\Z_d}\to\V$ is an embedding of $\Z_d$ into
 $\G^{(h)}$, then, $\forall u \in \V, \;\;\exists \bar{u} \in
 \eta(\V_{\Z_d})\subseteq \V$ such that $d_\G(u,\bar{u})\leq c$.
 \end{enumerate}
 Moreover, if $f:\V\to\R^d$ is a dense drawing of $\G$ in $\R^d$, then the embedding function $\eta$ in (ii) can be chosen so that $\forall u,v \in\V$, we can find $ u_z,v_z\in\V_{\Z_d}$ satisfying
\begin{align}\label{eq:dense-embedding-distancerelation}
\begin{split}
d_{\G}(u,\eta(u_z)) & \leq c, \quad d_{\G}(v,\eta(v_z)) \leq c \\
d_{\Z_d}(u_z,v_z)  & \leq 4d + \frac{\sqrt{d}}{\gamma}d_{f}(u,v)
\end{split}
\end{align}
where $\gamma$ is the maximum uncovered diameter of the $f$-drawing of $\G$. \frqed
\end{theorem}

In words, the two conditions state that $\G$ is dense in $\R^d$ if and only if (i) the
  $d$-dimensional lattice can be embedded in an $h$-fuzz of $\G$ for
  some positive integer $h$ and (ii) every node of $\G$ that is not
  the image of a node in $\Z_d$ is at a uniformly bounded graphical
  distance from a node that is the image of a node in $\Z_d$.  The
  significance of~\eqref{eq:dense-embedding-distancerelation} is that
  not only we can find for every node in $\G$ a close-by node that has
  a pre-image in the lattice, but also these close-by nodes can be so
  chosen so that if the Euclidean distance between a pair of nodes $u$
  and $v$ in the drawing is small, then the graphical
  distance in the lattice between the pre-images of their close-by
  nodes is small as well.

\medskip

\section{Electrical Analogy}\label{sec:analogy}
A crucial step in proving the main results of this paper is the
analogy  introduced in~\cite{PB_JH_TSP:08} between the BLU estimation problem and an abstract electrical network, where currents, potentials and resistances are matrix valued.

\medskip

A \emph{generalized electrical network} $(\G,R)$ consists of a graph
$\G=(\mbf V,\mbf E)$ (finite or infinite) together with a function
$R:\E\to \S^{k+}$ that assigns to each edge $e\in\E$ a
symmetric positive definite matrix $R_e$ called the \emph{generalized
resistance} of the edge. 

\medskip

A \emph{generalized flow from node $u\in\V$ to node $v \in \V$ with
  intensity $\mbf{j}\in\R^{k\times k}$} is an edge-function
  $j:\E \to \R^{k \times k}$ such that
\begin{align}\label{eq:KCL}
\sum_{\substack{(p,q)\in\mbf        E\\p=\bar        p}}       j_{p,q}
-\sum_{\substack{(q,p)\in\mbf  E\\p=\bar  p}} j_{q,p}  &=\begin{cases}
\mbf{j}  & \bar  p=u\\  -\mbf{j} &  \bar  p=v\\ \mbf{0}  &
\text{otherwise}
\end{cases} \forall \bar p \in \V.
\end{align}
A flow $j$ is said to have finite support if
it is zero on all but a finite number of edges. We say that a
flow $i$ is a \emph{generalized current} when there exists a
\emph{node-function} $\volt :\mbf V\to\R^{k\times k}$ for which
\begin{align}\label{eq:Ohmslaw}
  R_{u,v} i_{u,v}= \volt_u - \volt_v, \quad \forall (u,v)\in \E.
\end{align}
The node-function $\volt$ is called a \emph{generalized potential
  associated with the current $i$}. Eq.~\eqref{eq:KCL} should be
  viewed as a generalized version of Kirchhoff's current law and can be
  interpreted as: the net flow out of each node other than $u$ and $v$
  is equal to zero, whereas the net flow out of $u$ is equal to the
  net flow into $v$ and both are equal to the flow intensity
  $\mbf{j}$. Eq.~\eqref{eq:Ohmslaw} provides in a combined
  manner, a generalized version of Kirchhoff's loop law, which states
  that the net potential drop along a circuit must be zero, and Ohm's
  law, which states that the potential drop across an edge must be
  equal to the product of its resistance and the current flowing
  through it. A circuit is an undirected path that starts and ends at
  the same node. For $k=1$, generalized electrical networks are the
  usual electrical networks with scalar currents, potentials, and
  resistors.

\subsection{Effective Resistance and BLUE Error Covariance}

It was shown in~\cite{PB_JH_TSP:08} that when a
current of intensity $\i\in\R^{k \times k}$ flows from node
$u$ to node $v$, the resulting generalized current $i$ is a linear
function of the intensity $\i$ and there exists a matrix
$R^\mrm{eff}_{u,v}\in\S^{k+}$ such that
\begin{align}\label{eq:Reff_def}
V_u-V_v=R^\mrm{eff}_{u,v}\i, \quad \forall
\i\in\R^{k\times k}.
\end{align}
  We call the matrix $R^\mrm{eff}_{u,v}$ the \emph{generalized
  effective resistance between $u$ and $v$}.  In view of this
  definition, the effective resistance between two nodes is the
  generalized potential difference between them when a current with
  intensity equal to the identity matrix $I_k$ is injected at one node and extracted at the other, which is
  analogous to the definition of effective resistance in scalar
  networks~\cite{DoyleSnell}. Note that the effective resistance
  between two arbitrary nodes in a generalized network is a symmetric
  positive definite matrix as long as the network satisfies
  Assumption~\ref{as:bounded}, whether the network is finite or
  infinite~\cite{PB_JH_TSP:08}.

\medskip

Generalized electrical networks are useful in studying the BLU
estimation error in large networks because of the following analogy
between the BLU estimation error covariance and the generalized
effective resistance. 

\smallskip

\begin{theorem}[Electrical Analogy, from~\cite{PB_JH_TSP:08}]\label{thm:analogy}
Consider a measurement network $(\G,P)$ satisfying
Assumption~\ref{as:bounded} with $\G=(\V,\E)$ and a single reference
node $o\in\V$. Then, for every node $u\in \V \setminus \{o\}$, the
BLUE error covariance $\Sigma_{u,o}$ defined in~\eqref{eq:Sigma-infinite-defn} is a symmetric positive definite matrix equal to the generalized effective resistance $R_{u,o}^{eff}$ between $u$ and $o$ in the generalized electrical network $(\G, P)$:
\begin{align*}
  \Sigma_{u,o} = R^\mrm{eff}_{u,o}. \tag*{\frqed}
\end{align*}
\end{theorem}

\smallskip



\begin{remark}\label{rem:parallel}
In an electrical network, parallel resistors can be combined into one resistor by using the parallel resistance formula so that the effective resistance between every pair of nodes in the network remain unchanged. The same can be done in generalized electrical networks~\cite{BarooahThesis:07}. The analogy between BLUE covariance and effective resistance means that parallel measurement edges with possibly distinct measurement error covariances can be replaced by a single edge with an equivalent error covariance, so that the BLUE error covariances of all nodes remain unchanged. This explains why the assumption of not having parallel edges made at the beginning is not restrictive in any way. 
\end{remark}

\subsection{Graph Embedding and Partial Ordering of BLUE Covariances}\label{sec:embedding-orderingcovars}
Effective resistance in scalar electrical networks satisfies Rayleigh's Monotonicity Law, which states that the effective
resistance between any two nodes can only increase if the resistance
on any edge is increased, and vice versa~\cite{DoyleSnell}. The next
result (proved in~\cite{PB_JH_TSP:08}), states that the same is true for generalized networks, whether finite or infinite. 

\smallskip

\begin{theorem}[Rayleigh's Monotonicity Law~\cite{PB_JH_TSP:08}]\label{thm:Rayleigh}
 Consider two generalized electrical networks $(\G,R)$ and
  $(\bar{\G},\bar{R})$ with graphs $\G=(\V,\E)$ and
  $\bar\G=(\bar\V,\bar\E)$, respectively, such that both the networks satisfy 
  Assumption~\ref{as:bounded}. Assume that
  \begin{enumerate}
  \item $\G$ can be embedded in $\bar \G$, i.e.,  $\G\subset \bar\G$, and
  \item $R_e\geq\bar R_{\bar e}$ for every edge $e\in\E$.
  \end{enumerate}
  Then, for every pair of nodes $u,v\in\V$ of $\G$, 
  \begin{align*}
    R^\mrm{eff}_{u,v}\geq \bar{R}^\mrm{eff}_{u,v}
  \end{align*}
  where $R^\mrm{eff}_{u,v}$ and $\bar{R}^\mrm{eff}_{u, v}$
  are the effective resistance between $u$ and $v$ in the networks
  $(\G,R)$ and $(\bar{\G}, \bar{R})$, respectively. \frqed
\end{theorem}

\medskip

The usefulness of Rayleigh's Monotonicity Law in answering the error
scaling question becomes apparent when combined with the Electrical
Analogy. It shows that when $\G$ can be embedded in $\bar{\G}$, the
BLUE error covariances in $\G$ are lower bounded by the error covariances in
$\bar{\G}$. Intuitively, since $\G$ has only a subset of the
measurements in $\bar{\G}$, the estimates in $\G$ are less
accurate than those in $\bar{\G}$.

\medskip

\begin{remark}\label{rem:Reff-directions}
Although the graph $\G$ that defines the electrical network $(\G,R)$
  is directed, the edge directions are irrelevant in determining
  effective resistances. This is why Rayleigh's Monotonicity Law holds
  with graph embedding, which is insensitive to edge directions. The
  electrical analogy also explains why the edge directions are
  irrelevant in determining error covariances.\frqed
\end{remark}

\subsection{Triangle Inequality}
Matrix-valued effective resistances satisfy a triangle inequality, which will be useful in proving the error scaling laws in Section~\ref{sec:mainproof}. It is known that scalar effective resistance obeys triangle inequality, and is therefore also referred
to as the ``resistance distance''~\cite{DJK_MR_JMC:93}.  Although the
result in~\cite{DJK_MR_JMC:93} was proved only for finite networks, it
is not hard to extend it to infinite networks.  The following
simple extension of the triangle inequality to generalized networks
with constant resistances on every edge was derived in~\cite{BarooahThesis:07}:

\medskip

\begin{lemma}[Triangle Inequality]\label{lem:triangle} 
Let $(\G,R_o)$ be a generalized electrical network satisfying
Assumption~\ref{as:bounded} with a constant resistance $R_o \in
\S^{k+}$ on every edge of $\G$. Then, for every triple of nodes
$u,v,w$ in the network,
\begin{align*}
R^\mrm{eff}_{u,w} \leq R^\mrm{eff}_{u,v} + R^\mrm{eff}_{v,w}. \tag*{\frqed}
\end{align*}
\end{lemma}

\subsection{Effective Resistances in Lattices and Fuzzes}\label{sec:Reff-specialgraphs}
Recall that given a graph $\G$ and a positive integer $h$, the $h$-fuzz of
$\G$, denoted by $\G^{(h)}$, is a graph that has an edge between two
nodes $u$ and $v$ whenever the graphical distance between them in $\G$
is less than or equal to $h$. 

\smallskip

An $h$-fuzz will clearly have lower effective
resistance than the original graph because of Rayleigh's Monotonicity
Law, but it is lower only by a constant factor as stated in the following result, which is a straightforward extension to the generalized case of a result about scalar effective resistance established by Doyle and Snell (see the Theorem on page $103$, as well as Exercise $2.4.9$, in~\cite{DoyleSnell}). The interested reader can find a proof in~\cite{BarooahThesis:07}.

\medskip

\begin{lemma}\label{lem:fuzzing-effres} Let $(\G,R_o)$ be a generalized
electrical network satisfying Assumption~\ref{as:bounded} with a
constant generalized resistance $R_o \in \S^{k+}$ on its
every edge. Let $(\G^{(h)},R_o)$ be the electrical network similarly
constructed on $G^{(h)}$, the $h$-fuzz of $G$.  For every pair of
nodes $u$ and $v$ in $\V$,
\begin{align*}
\alpha R^\mrm{eff}_{u,v}(\G) & \leq R^\mrm{eff}_{u,v}(\G^{(h)}) \leq R^\mrm{eff}_{u,v}(\G),
\end{align*}
where $R^\mrm{eff}_{u,v}(\cdot)$ is the effective resistance in the
network $(\cdot,R_o)$ and $\alpha \in (0,1]$ is a positive constant that
does not depend on $u$ and $v$. \frqed
\end{lemma}

\medskip

The following lemma establishes effective resistances in
$d$-dimensional lattices and their fuzzes. 

\medskip

\begin{lemma}\label{lem:lattice-fuzz-effres}
For a given positive integer $h$, consider the electrical network $(\Z_d^{(h)},R_o)$ with a constant
generalized resistance $R_o \in \S^{k+}$ at every edge of the $h$-fuzz
of the $d$-dimensional square lattice $\Z_d$. The generalized effective resistance $R^{\mrm{eff}}_{u,v}$
between two nodes $u$ and $v$ in the electrical network $(\Z_d^{(h)},R_o)$
satisfies
\begin{enumerate}
\item{$ R^{\mrm{eff}}_{u,v}(\Z^{(h)}_1) =
 \Theta\big( d_{\Z_1}(u,v)\big) $}
 \item{$ R^{\mrm{eff}}_{u,v}(\Z^{(h)}_2) = \Theta
 \big(  \log d_{\Z_2}(u,v) \big) $, }
\item{$R^{\mrm{eff}}_{u,v}(\Z^{(h)}_3) = \Theta \big( 1 \big) $. \frqed}
\end{enumerate}
\end{lemma}
\begin{proof-lemma}{\ref{lem:lattice-fuzz-effres}}
The \emph{scalar} effective resistance in $1$D , $2$D, and $3$D lattices follow linear, logarithmic and bounded growth rates, respectively~\cite{Cserti2000,ResistiveLattice_Atkinson99}. Using these results, it was established in~\cite{PB_JH_TSP:08} that the \emph{matrix} effective resistances in these lattices have the same scaling laws (see Lemma $5$ in~\cite{PB_JH_TSP:08}).
Thus, 1D, 2D, and 3D lattices with matrix-valued resistances have linear, logarithmic, and bounded scaling laws for the effective resistance, which is the result with $h=1$. The case $h>1$
follows from the application of Lemma~\ref{lem:fuzzing-effres}.
\end{proof-lemma}

The slowing down of the growth of the effective resistance as the dimension increases can be attributed to the fact that the number of paths between each pair of nodes is larger in higher dimensional lattices.  The scaling laws for effective resistance in lattices and their fuzzes also have intimate connections to the change from recurrence to transience of random walks in lattices as the dimension changes from $d=1,2$ to $d \geq 3$~\cite{DoyleSnell}.

%
%
\section{Proof of Theorem~\ref{thm:errorscalinglaws}}\label{sec:mainproof}
We now prove Theorem~\ref{thm:errorscalinglaws} by using the tools
that have been developed so far.  The following terminology is 
needed for the proofs. For functions $g: \R \to \R^{k \times k}$ and $p: \R \to \R$, the  notation $g(y)=\Theta(p(y))$ means that $g(y) = \Omega(p(y))$ and
$g(y) = \scr{O}(p(y))$.  The notations $\scr{O}(\cdot)$ and
$\Omega (\cdot)$ are described in Section~\ref{sec:mainresults}.

\begin{proof-theorem}{\ref{thm:errorscalinglaws}}
[Upper bounds:] 
We start by establishing the upper bounds on the effective resistance for graphs that are dense in $\R^d$.  Throughout the proof of the upper bounds, we will use
$R^\mrm{eff}_{u,v}(\G)$, for any graph $\G$, to denote the effective
resistance between nodes $u$ and $v$ in the electrical network
$(\G,P_{\max})$ with every edge of $\G$ having a generalized
resistance of $P_{\max}$.  
From the Electrical Analogy theorem and Monotonicity Law (Theorems
\ref{thm:analogy} and \ref{thm:Rayleigh}), we get
\begin{equation*}
\Sigma_{u,o} \leq R^\mrm{eff}_{u,o}(\G).
\end{equation*}
To establish an upper bound on $\Sigma_{u,o}$, we will now establish an upper bound on the resistance  $R^\mrm{eff}_{u,o}(\G)$.  To this effect, suppose that $f$ is a dense drawing of $G$ in $\R^d$. From dense embedding Theorem~\ref{thm:embedding-dense}, we conclude that there exists a positive integer $h$ such that the $d$-D lattice $\Z_d$ can be embedded in the $h$-fuzz of $\G$. Moreover, Theorem~\ref{thm:embedding-dense} tells us that there exists 
$u_z,o_z\in\V_{\Z_d}$, a positive constant $c$, and an embedding
$\eta:\V_{\Z_d}\to\V$ of $\Z_d$ into $\G^{(h)}$, such that
\begin{align}
 & d_{\G}(u,\eta(u_z))\leq c, \; d_{\G}(o,\eta(o_z))\leq c \label{eq:df-dz-1} \\
 & d_{\Z_d}(u_z,o_z) < 4d + \frac{\sqrt{d}}{\gamma}d_f(u,o), \label{eq:df-dz-2}
\end{align}
where $\gamma$ is the maximum uncovered diameter of the $f$-drawing of
$\G$. Note that $\eta(u_z),\eta(o_z) \in \V$. 
Consider the electrical network $(\G^{(h)},P_{\max})$ formed by
assigning to every edge of $G^{(h)}$ a resistance of $P_{\max}$.  From
the triangle inequality for effective resistances
(Lemma~\ref{lem:triangle}),
\begin{align}\label{eq:Reff-triangle-denseproof}
R^\mrm{eff}_{u,o}(\G^{(h)}) & \leq R^\mrm{eff}_{u,\eta(u_z)}(\G^{(h)})+R^\mrm{eff}_{\eta(u_z),\eta(o_z)}(\G^{(h)})\nonumber \\
&\qquad +R^\mrm{eff}_{\eta(o_z),o}(\G^{(h)}).
\end{align}
For any two nodes $u,v\in\V$, application of the triangle inequality Lemma~\ref{lem:triangle} to successive nodes on the shortest path joining $u$ and $v$ gives us
$R^\mrm{eff}_{u,v}(\G^{(h)})\leq d_{\G^{(h)}}(u,v)P_{\max} \leq
\frac{2}{h}d_{\G}(u,v)P_{\max}$. Using this bound in~\eqref{eq:Reff-triangle-denseproof}, and by using~\eqref{eq:df-dz-1}, we conclude that
\begin{align}\label{eq:uubar-dense}
R^\mrm{eff}_{u,o}(\G^{(h)}) \leq \frac{4c}{h}P_{\max}+R^\mrm{eff}_{\eta(u_z),\eta(o_z)}(\G^{(h)}).
\end{align}
Since $\G^{(h)} \supset \Z_d$, from Rayleigh's Monotonicity Law
(Theorem~\ref{thm:Rayleigh}), we obtain
\begin{align*}
R^\mrm{eff}_{\eta(u_z),\eta(o_z)}(\G^{(h)}) & \leq  R^\mrm{eff}_{u_z,o_z}(\Z_d).
\end{align*}
When $\G$ is dense in, say, in $\R^2$, we have from
Lemma~\ref{lem:lattice-fuzz-effres} that
\[
R^\mrm{eff}_{u_z,o_z}(\Z_2) = \Theta\left(\log d_{\Z_2}(u_z,o_z) \right),
\]
which implies 
\begin{align*}
R^\mrm{eff}_{\eta(u_z),\eta(o_z)}(\G^{(h)}) & = \scr{O}\left(\log d_{\Z_2}(u_z,o_z) \right).
\end{align*}
Combining this with~\eqref{eq:df-dz-2} and \eqref{eq:uubar-dense}, we get
\[
R^\mrm{eff}_{u,o}(\G^{(h)}) = \scr{O}\left(\log d_{f}(u,o) \right).
\]
From Lemma~\ref{lem:fuzzing-effres} we know that the effective resistance
in $\G$ and its $h$-fuzz is of the same order, so that 
\[
R^\mrm{eff}_{u,o}(\G) = \Theta\left(R^\mrm{eff}_{u,o}(\G^{(h)})\right),
\]
from which the desired result follows:
\[
\Sigma_{u,o} \leq R^\mrm{eff}_{u,o}(\G) = \scr{O}\left(\log d_{f}(u,o) \right).
\]
The statements of the upper bounds for $1$ and $3$-dimensions can be
proved similarly. This concludes the proof of the upper bounds in Theorem~\ref{thm:errorscalinglaws}.

\medskip

[Lower bounds:]  Now we establish the lower bounds on the BLUE error covariance $\Sigma_{u,o}$ in a sparse graph. Throughout the proof of the lower bounds, for a graph $\G$, we will use $R^\mrm{eff}_{u,v}(\G)$ to denote the
effective resistance between nodes $u$ and $v$ in the electrical
network $(\G,P_{\min})$ with every edge of $\G$ having a generalized
resistance of $P_{\min}$. From
the Electrical Analogy and Rayleigh's Monotonicity Law
(Theorems~\ref{thm:analogy} and~\ref{thm:Rayleigh}), we get
\begin{equation}\label{eq:analogy_lowerbnd-1}
\Sigma_{u,o} \geq R^\mrm{eff}_{u,o}(\G).
\end{equation}
Therefore, to establish a lower bound on $\Sigma_{u,o}$, we proceed by establishing a lower bound on the resistance $R^\mrm{eff}_{u,o}(\G)$. Since $\G$ is sparse in $\R^d$, it follows from
Theorem~\ref{thm:embedding-sparse} that there exists a positive
integer $h$, such that $\G \subset \Z_d^{(h)}$.  Let $\eta: \V \to
\V_{\Z_d}$ be the embedding of $\G$ into $\Z_d^{(h)}$.  Consider the
generalized electrical network $(\Z_d^{(h)},P_{\min})$ formed by
assigning a generalized resistance of $P_{\min}$ to every edge of
$\Z_d^{(h)}$. From Rayleigh's monotonicity law, we get
\begin{align}\label{eq:analogy_lowerbnd-4}
R^\mrm{eff}_{u,o}(\G) \geq R^\mrm{eff}_{u_z,o_z}(\Z_d^{(h)}),
\end{align}
where $u_z=\eta(u),o_z=\eta(o)$ refer to the nodes in $\Z_d^{(h)}$
that correspond to the nodes $u,o$ in $\G$. When the graph is sparse
in, say, $\R^2$, it follows from~\eqref{eq:analogy_lowerbnd-4} and Lemma~\ref{lem:lattice-fuzz-effres} that
\begin{align*}
R^\mrm{eff}_{u,o}(\G) & = \Omega\left(\log d_{\Z_2}(u_z,o_z)\right)\\
            & = \Omega\left(\log d_{f}(u,o)\right),
\end{align*}
where the second statement follows
from~\eqref{eq:civil-embedding-distancerelation} in
Theorem~\ref{thm:embedding-sparse}. Combining the above
with~\eqref{eq:analogy_lowerbnd-1}, we get $\Sigma_{u,o}= \Omega(\log
d_f(u,o))$, which proves the lower bound for graph that are sparse in $\R^2$. The statements for the lower bounds graphs that are sparse in $\R^1$ or $\R^3$ can be proved in an analogous manner. This concludes the proof of the theorem. 
\end{proof-theorem}

\section{Checking Denseness and Sparseness}\label{sec:checking-denseness}
To show that a graph is dense (or sparse) in a particular dimension, one
has to find a drawing in that dimension with the appropriate
properties. For sensor networks, sometimes the \emph{natural drawing} of a
deployed network is sufficient for this purpose. By the natural drawing of
a sensor network we mean the mapping from the nodes to their physical locations
in the Euclidean space in which they are deployed.  We can use this 
natural drawing to construct the following examples of dense and sparse graphs.

\begin{proposition}\label{prop:examples}
\begin{enumerate}
\item 
Deploy a countable number of nodes in $\R^d$ so that the maximum uncovered diameter $\gamma$ of its natural drawing is finite, and allow every pair of nodes whose Euclidean distance is no larger than $2\gamma$ to have an edge between them. The resulting graph is weakly connected and dense in $\R^d$. Such a graph is also sparse in $\R^d$ if the nodes are placed such that every finite volume in $\R^d$ contains a finite number of nodes. 
\item
Consider an initial deployment of nodes on a square lattice in $\R^2$, for which a fraction of the nodes has subsequently failed. Suppose that the number of nodes that failed in any  given region is bounded by a linear function of the area of the region, i.e., that there exist constants $\alpha$ and $\beta$ such that, for every region of area $A$ the number of nodes that failed in that region is no larger than $\alpha A + \beta$. Assuming that $\alpha < \frac{1}{4(\beta+1)}$, there will be an infinite connected component among the remaining nodes, which is dense and sparse in $2$-D. \frqed
\end{enumerate}
\end{proposition}

\medskip

\ifthenelse{\equal{\PaperORReport}{Report}}{
The proof of the proposition above is provided in Appendix~\ref{sec:App-proofs-1}.
}
{
The interested reader may consult~\cite{PB_JH_scaling_ArXiV:08} for the proof of the proposition.
} 

\medskip

The first example in the proposition is that of a geometric graph that is obtained by placing a number of nodes in a region and specifying a range such that a pair of nodes have an edge between them if and only if the Euclidean distance between them is no more than the given range. The second example refers to a network in which some of the initially deployed nodes have failed, with the stipulation that in large areas, no more than a certain fraction of the node may fail. 
For example, $\beta= 5$ and $\alpha = 0.04$ satisfies the stated  conditions. It can be shown that $\beta=5$ and $\alpha=0.04$ means that in areas larger than $10 \times 10$, at most $4\%$ of the nodes may fail.

\medskip

To show that a graph is not dense (or not sparse) in a particular dimension is harder since one has to show that no drawing with the required properties exists. Typically, this can be
done by showing that the existence of a dense (or sparse) drawing leads to a contradiction. An application of this technique leads to the following result.

\begin{lemma}\label{lem:lattice-dense-sparse}
\begin{enumerate}
\item The $d$-dimensional lattice $\Z_d$ is not sparse in
$\R^{\underline d}$ for every $\underline{d} < d$, and it is not dense
in $\R^{\overline d}$ for every $\overline{d} > d$.
\item A regular-degree\footnote{A graph is called regular-degree if
the degree of every node in the graph is the same. } infinite tree is
not dense or sparse in any dimension. \frqed
\end{enumerate}
\end{lemma}

\ifthenelse{\equal{\PaperORReport}{Report}}{
The first statement of the lemma is provided in Appendix~\ref{sec:App-proofs-1}. The
proof of the second statement is not provided since the method of the
proof is similar.
}
{
The proof of the first statement of the lemma is provided in the Appendix. The second statement can be proved in an analogous manner.
}

\smallskip

We are now ready to prove Lemma~\ref{lem:noconfusion}.

\smallskip

\begin{proof-lemma}{\ref{lem:noconfusion}}
To prove the result by contradiction, suppose that a graph $\G$ is
dense in $\R^d$ as well as sparse in $\R^{d'}$, where $d'<d$. It follows from Theorems~\ref{thm:embedding-dense}
and~\ref{thm:embedding-sparse} that there exist positive integers
$\ell,p$ such that $\Z_d \subset \G^{(\ell)}$ and $\G \subset \Z_{d'}^{(p)}$. It is straightforward to verify the following facts:
\begin{enumerate}
\item for every pair of
graphs $\G,\bar\G$ that do not have any parallel edges, $\G
\subset \bar\G \Rightarrow \G^{(l)} \subset {\bar\G}^{(l)}$ for every
positive integer $l$. 
\item for an arbitrary graph $\G$ without parallel edges, and
two positive integers $\ell,p$, we have $(\G^{(p)})^{(\ell)} =
\G^{(p\ell)}$. 
\end{enumerate}
It follows that $\Z_d \subset \Z_{d'}^{(\ell p)}$, which means,
from sparse embedding Theorem~\ref{thm:embedding-sparse}, that a
$d$-dimensional lattice is sparse in $\R^{d'}$. This is a
contradiction because of Lemma~\ref{lem:lattice-dense-sparse}, which
completes the proof.
\end{proof-lemma}

\section{Summary and Future Work}\label{sec:summary}
In a large number of sensor and ad-hoc network applications, a number
of node variables need to be estimated from measurements of the noisy
differences between them. This estimation problem is naturally posed in terms of a graph.

\medskip

We established a classification of graphs, namely, dense or sparse in $\R^d,\;1\leq d \leq 3$, that determines how the optimal linear unbiased estimation 
error of a node grows with its distance from the reference
node. The notion of denseness/sparseness introduced in this paper is distinct from the usual notion based on the average degree. In fact, we illustrated through examples that node degree is a poor measure of how the estimation error scales with distance. 

\medskip

The bounds and the associated graph classification derived here can be used in performance analysis, design and deployment of large networks. For example, if a sensor network is sparse in $\R$, then we know that the estimation error of a node will grow linearly with its distance from a reference. A large number of reference nodes will thus be needed for large networks that are sparse in $\R$. On the other hand, if one has control over the network deployment, then one should strive to obtain a network that is dense in $\R^d$ with $d$ as large as possible. In the ideal case of $d=3$, with a single reference node one can get bounded estimation error regardless of how large the network is. 

\medskip

There are several avenues for future research. The scaling laws described in this paper were derived for infinite measurement graphs. This is justified by the fact that the BLUE covariance $\Sigma_{u,o}$ of a  node $u$ in an infinite graph is very close to the obtained in a large finite subgraph that contains the nodes $u$ and $o$ sufficiently inside it~\cite{PB_JH_TSP:08}. However, to gain a better understanding of the ``boundary'' effects that can occur in finite graphs, an interesting research direction is to determine how large the BLUE error covariance can be as a function of the size of the graph, for nodes that are close to the edge of the graph. A connection between the notions introduced in this paper and those in coarse geometry might be useful in this regard.  It can be shown that a graph $\G$ that is both sparse and dense in $\R^d$ is \emph{coarsely equivalent} to $\R^d$, which intuitively means that $\G$ and $\R^d$ are the same in their large scale structure (see~\cite{coarsegeom_JR:03} for a precise definition of coarse equivalence). Certain coarse geometric notions that were originally defined for infinite graphs have been extended to finite graphs (see~\cite{EJ_PL_FB_JGT:08}). This connection between coarse geometry and denseness/sparseness might provide a way to extend the techniques used in this paper to finite graphs.

\medskip

Although the dense and sparse classification does allow randomness in the structure of the graph, the effect of such randomness on the scaling laws for the error is not explicitly accounted for in the present work. A useful research
direction would be to characterize the estimation error
covariances in graphs with random structure, such as random geometric
graphs~\cite{RGG_Penrose:03}. Another interesting avenue for future research is the investigation of estimation error growth in scale-free networks that do not
satisfy the bounded degree assumption.

\bibliographystyle{abbrvunsrt}
\bibliography{../../../PBbib/sensnet_bib_dbase}

\appendices


\section{Technical Proofs}\label{sec:App-proofs-1}
\begin{proof-lemma}{\ref{lem:rho}}
  We prove that \ref{en:rho} implies \ref{en:alpha-beta} by
  contradiction. Assuming that \ref{en:alpha-beta} does not hold, we
  have that
  \begin{align*}
    \forall \alpha>0 \quad\forall \beta>0 \quad \exists \bar u,\bar v\in\V \text{ such that }
    d_\G(\bar u,\bar v)> \alpha d_f(\bar u,\bar v)+\beta.
  \end{align*}
  or equivalently
  \begin{align*}
    \forall \alpha>0 \quad\forall \beta>0 \quad \exists \bar u,\bar v\in\V
    \end{align*}
    such that
    \begin{align*}
    \frac{d_f(\bar u,\bar v)}{d_\G(\bar u,\bar v)}<\frac{1}{\alpha}-\frac{\beta}{\alpha d_\G(\bar u,\bar v)}.
  \end{align*}
  This means that for a given $\alpha>0,\beta>0$, the set
  \begin{align*}
    \Big\{\frac{d_f(u,v)}{d_\G(u,v)}: u,v\in\V \text{ and } d_\G(u,v)\geq \beta\Big\}
  \end{align*}
  contains at least the element
  \begin{align*}
    \frac{d_f(\bar u,\bar v)}{d_\G(\bar u,\bar v)}
    <\frac{1}{\alpha}-\frac{\beta}{\alpha d_\G(\bar u,\bar v)}< \frac{1}{\alpha}
  \end{align*}
  and therefore
  \begin{align*}
    \inf \Big\{\frac{d_f(u,v)}{d_\G(u,v)}: u,v\in\V
      \text{ and } d_\G(u,v)\geq \beta \Big\}<\frac{1}{\alpha}.
  \end{align*}
  Making $\beta\to\infty$ we obtain that
  \begin{align*}
    \rho = \lim_{\beta\to \infty} \inf \Big\{\frac{d_f(u,v)}{d_\G(u,v)}: u,v\in\V
      \text{ and } d_\G(u,v)\geq \beta\Big\}<\frac{1}{\alpha}.
  \end{align*}
  But since $\alpha$ can be arbitrarily large, the above actually implies
  that $\rho=0$, which contradicts \ref{en:rho}.

  \medskip

  To prove that \ref{en:alpha-beta} implies \ref{en:rho}, we note that
  when \ref{en:alpha-beta} holds, we conclude that for every pair of
  nodes $u,v\in\V$, for which $d_\G(u,v)\geq n$, we have that
  \begin{align*}
    \frac{d_f(u,v)}{d_\G(u,v)}\geq \frac{1}{\alpha}-\frac{\beta}{d_\G(u,v)}
    \geq\frac{1}{\alpha}-\frac{\beta}{n} ,
    \quad \forall u\not= v \in \V.
  \end{align*}
  Therefore,
  \begin{align*}
    \inf \Big\{\frac{d_f(u,v)}{d_\G(u,v)}: u,v\in\V \text{ and } d_\G(u,v)\geq n\Big\}
    \geq \frac{1}{\alpha}-\frac{\beta}{n}.
  \end{align*}
  As $n\to\infty$, the left-hand side converges to $\rho$ and the right-hand
  side converges to $\frac{1}{\alpha}>0$, from which \ref{en:rho} follows.
\end{proof-lemma}

\medskip

\begin{proof-theorem}{\ref{thm:embedding-dense}}
In this proof, we will denote by $g:\V_{\Z_d}\to\R^d$ the natural drawing of the lattice $\Z_d$.

($\Rightarrow$) We have to prove that if $\G$ is dense in $\R^d$,
conditions (i) and (ii) are satisfied. Since $\G$ is dense in $\R^d$,
there is a drawing function $f:\V\to\R^d$ so that the $f$-drawing of
$\G$ has a $\gamma<\infty$ and $\rho>0$.  Define a new drawing
$f':\V\to\R^d$ as
\begin{align*}
f'(u)=\frac{1}{\gamma}f(u),\quad\forall u\in\V,
\end{align*}
so that the maximum uncovered diameter $\gamma'$ of the $f'$ drawing
of $\G$ is $1$.  Note that $f'$ is still a dense drawing of $\G$. Now
we superimpose the natural $g$-drawing of $\Z_d$ on the $f'$-drawing
of $\G$, and draw open balls of diameter $1$ centered at the natural drawing $g(u_z)$ of every lattice node, denoted by $B(g(u_z),\frac{1}{2})$.  Figure~\ref{fig:dense_embedding_proof} shows an example in
$\R^2$. Since $\gamma' =1$, it follows from the definition of
denseness that in every one of those balls, there is at least one node
$u \in \V$.
\begin{figure}[htb]
\psfrag{g}{$\gamma'=1$} \psfrag{u}{$\bar u$} \psfrag{v}{$\bar v$} \psfrag{uz}{$ u_z$}
\psfrag{vz}{$v_z$}
\psfrag{df}{}
\begin{center}
\includegraphics[scale=0.8]{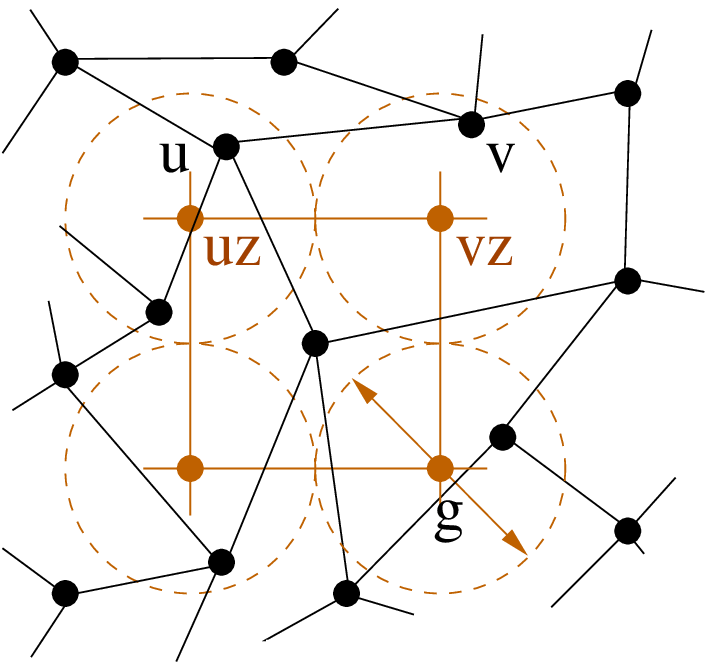}
\caption{\label{fig:dense_embedding_proof}Superimposing a 2-dimensional
  lattice (gray/brown) on a 2-dimensional dense graph (black).}
\end{center}
\end{figure}
To construct the embedding, we associate each node of the lattice to a
node of $\G$ whose drawing appears inside the ball centered around the
lattice node. This defines an injective function $\eta:\V_{\Z_d} \to
\V$. Consider two nodes of the lattice $u_z,v_z \in \V_{\Z_d}$ that
have an edge between them. Let $\bar u \eqdef \eta(u_z), \bar v \eqdef
\eta(v_z)$. Since $f'(\bar u)$ and $f'(\bar v)$ belong to adjacent
balls of unit diameter (see Figure~\ref{fig:dense_embedding_proof}),
\[
d_{f'}(\bar u,\bar v)=\|f'(\bar u)-f'(\bar v)\| \leq 2.
\]
Since $f'$ is a dense drawing in
$\R^d$ with $\gamma'=1$, it follows from Lemma~\ref{lem:rho} that $d_{\G}(\bar u,\bar v)\leq 2\alpha+\beta$, for
some positive constants $\alpha$ and $\beta$. Define $h \eqdef \lceil
2\alpha+\beta \rceil$. Then $\bar u$ and $\bar v$ will have an edge
between them in the $h$-fuzz $\G^{(h)}$. So $\G^{(h)} \supset \Z_d$,
and we have the desired result that denseness implies (i).

\medskip

To show that denseness implies (ii), first note that if
$u\in\eta(\V_{\Z_d})$, then (ii) is trivially true (choose $\bar u
\eqdef u$), so only nodes in $\V\setminus\eta(\V_{\Z_d})$ are
interesting. For every $u\in\V $, find
$u_z\in\V_{\Z_d}$ as the node in the lattice such that the ball of
unit diameter drawn around $u_z$ is closest to $u$. That is, find $u_z \in \V_{\Z_d}$ such that
\begin{align}\label{eq:finding-uz}
u_z = \arg \min_{u^{'}_z \in\V_{\Z_d} }\mrm{dist}\left(f'(u), B(g(u^{'}_z),1/2)\right)
\end{align}
where $\mrm{dist}(x,A)$ between a point $x\in\R^d$ and a set $A \subset \R^d$ is defined as 
\[
\mrm{dist}(x,A) = \inf_{y \in A} \|x - y\|.
\]
There are only $2^d$ balls one needs to check to determine the minimum
in~\eqref{eq:finding-uz}, so $u_z$ exists, though it may not be
unique. If there are multiple minima in~\eqref{eq:finding-uz}, pick any one. This procedure defines an onto map $\xi : \V \to
\V_{\Z_d}$. Let $\eta:\V_{\Z_d} \to \V$ be the embedding of $\Z_d$
into $\G^{(h)}$ as described earlier in this proof. Define $\psi:\V
\to \V$ as $\psi \eqdef (\eta \circ \xi)$. We will now show that, for
every $u\in\V$, the node $\psi(u)\in\V$, which has a corresponding
node in the lattice, is within a uniformly bounded graphical distance
of $u$. Since $f'(u)$ either lies in the ball centered at $g(u_z)$ or
in the gaps between that ball and the neighboring balls,
$\|f'(u)-g(u_z)\|< \sqrt{d}$. Therefore,
\begin{align}
d_{f'}(u,\psi(u))& \leq \|f'(u)-g(u_z)\|+\|g(u_z)-f'(\psi(u))\| \nonumber \\
                & < \sqrt{d} + \frac{1}{2} \leq \frac{3}{2}\sqrt{d}, \label{eq:df-u-ubar}
\end{align}
where we have used the fact that $f'(\psi(u)) \in
B(g(u_z),\frac{1}{2})$. From Lemma~\ref{lem:rho} and the denseness of the
$f$-drawing of $\G$, we get
\begin{align*}
d_{\G}(u,\psi(u)) & \leq \alpha d_f(u,\psi(u))+\beta \\
                  & = \alpha \gamma d_{f'}(u,\psi(u))+\beta \\
                  & < \frac{3}{2}\alpha \gamma \sqrt{d}+\beta.
\end{align*}
Define 
\begin{align}\label{eq:c-defn}
c\eqdef\lceil \frac{3}{2}\alpha \gamma \sqrt{d}+\beta \rceil,
\end{align}
 which is a constant independent of $u$ and $v$. Then for every
$u\in\V$, there exists a $\bar u \eqdef \psi(u)\in\eta(\V_{\Z_d})\subset \V$ such
that $d_{\G}(u,\bar u)<c$, which is the desired condition (ii).

\medskip

($\Leftarrow$) We have to prove that if (i) and (ii) are satisfied,
then $\G$ is dense in $\R^d$.  We will construct a drawing $f$ of $\G$
in $\R^d$ with the following procedure and then prove that it is a
dense drawing.  Since $\Z \subset \G^{(h)}$, there is an
injective map $\eta:\V_{\Z_d} \to \V$ such that $\eta(\V_{\Z_d})
\subset \V$. Pick a node $u$ in $\V$ that has not been drawn
yet. By (ii), there exists a positive constant $c$ and a node
$u_z\in\V_{\Z_d}$ such that $\bar{u} \eqdef \eta(u_z) \in \V$ and
$d_{\G}(u,\bar{u})<c$.  If $\bar u$ has not  been drawn yet, then draw
it the location of its corresponding lattice node, i.e.,
\begin{align}\label{eq:fu-guz}
f(\bar u) = g(u_z).
\end{align}
A little thought will reveal that if $\bar{u}$ has been drawn already,
as long as the drawing procedure outlined so far is followed, it must
have been drawn on the lattice location $g(u_z)$, so~\eqref{eq:fu-guz}
holds. Once $\bar u$ is drawn, we draw $u$ in the following way. In
case $\bar u = u$, drawing of $u$ is determined by the drawing of
$\bar u$. If $u \neq \bar{u}$, draw $u$ by choosing a random location
inside an open ball of diameter $1$ with the center at $f(\bar u)$. To
show that a drawing obtained this way is dense, first note that the
largest uncovered diameter $\gamma<2$ since a subset of the nodes of
$\V$ occupy the lattice node positions. Pick any two nodes $u,v\in\V$.
Again, from (ii), we know that there exists $\bar u,\bar v \in
\eta(\V_{\Z_d})\subset \V$ such that $d_\G(u,\bar u)\leq c$ and
$d_\G(v,\bar v)\leq c$ for some positive constant $c$. Therefore
\begin{align*}
d_\G(u,v) & \leq d_\G(u,\bar u) + d_\G(\bar u,\bar v) + d_\G(\bar v, v)\\
          & \leq 2c +  h\,d_{\G^{(h)}}(\bar u,\bar v)
\end{align*}
Since $\Z_d \subset \G^{(h)}$,
\begin{align*}
d_{\G^{(h)}}(\bar u,\bar v) & \leq d_{\Z_d}(\eta^{-1}(\bar u),\eta^{-1}(\bar v)) \\
                  & = \|g(u_z) - g(v_z)   \|_1 \\\intertext{where $\|\cdot\|_1$ denotes the vector $1$-norm,}
                  & \leq \sqrt{d}\|g(u_z) - g(v_z)   \| \\
                  & = \sqrt{d}\|f(\bar u) - f(\bar v))\| \text{ (from~\eqref{eq:fu-guz})} \\
                  & = \sqrt{d}\,d_f(\bar u,\bar v).
\end{align*}
Because of the way the drawing $f$ is constructed, we have
$d_f(u,\bar{u})\leq 1$, which implies $d_f(\bar u, \bar u) \leq
d_f(\bar u, u)+ d_f(u, v)+ d_f(v, \bar v) = d_f(u, v)+2$. So we have
\begin{align*}
d_\G(u,v) & \leq 2c+h \sqrt{d}\,(d_f(u,v)+2)\\
          & = 2(c+h\sqrt{d}) + h\sqrt{d}\; d_f(u,v).
\end{align*}
From Lemma~\ref{lem:rho}, we
see that the asymptotic distance ratio $\rho >0$ for the $f$-drawing
of $\G$, which establishes that $f$ is a dense drawing of $\G$ in $\R^d$.
It follows that $\G$ is dense in $\R^d$.

\medskip

To prove the relationship~\eqref{eq:dense-embedding-distancerelation}
for any dense drawing $f$, consider again the scaled drawing $f'$
defined as $f'=f/\gamma$, so that the maximum uncovered diameter of
$f'$ is $1$. Since $\G$ is dense in $\R^d$, $\Z_d$ can be embedded in
$\G^{(h)}$ with an embedding $\eta:\V_{\Z_d}\to\V$. We choose the
embedding $\eta$ as described in the first part of the proof. For
every $u\in\V$, call $u_z \eqdef \xi(u)$, where $\xi:\V \to \V_{\Z_d}$ was  defined earlier in this proof for the $f'$
dense drawing of $\G$. Now consider two arbitrary nodes $u,v\in\V$ and
let $u_z\eqdef \xi(u)$, $v_z \eqdef \xi(v)$ (see
Figure~\ref{fig:dense_embedding_part2}). It was shown earlier in this
proof that for every pair of nodes $u,v\in\V$, we have
$d_\G(u,\eta(u_z))<c$ and $d_\G(v,\eta(v_z))<c$, where $c$ is defined
in~\eqref{eq:c-defn}.

\begin{figure}[htb]
\psfrag{g}{$\gamma'=1$}
\psfrag{u}{$u=\bar u$}
\psfrag{v}{$v$}
\psfrag{vbar}{$\bar v$}
\psfrag{uz}{$u_z$}
\psfrag{vz}{$v_z$}
\begin{center}
\includegraphics[scale=0.6]{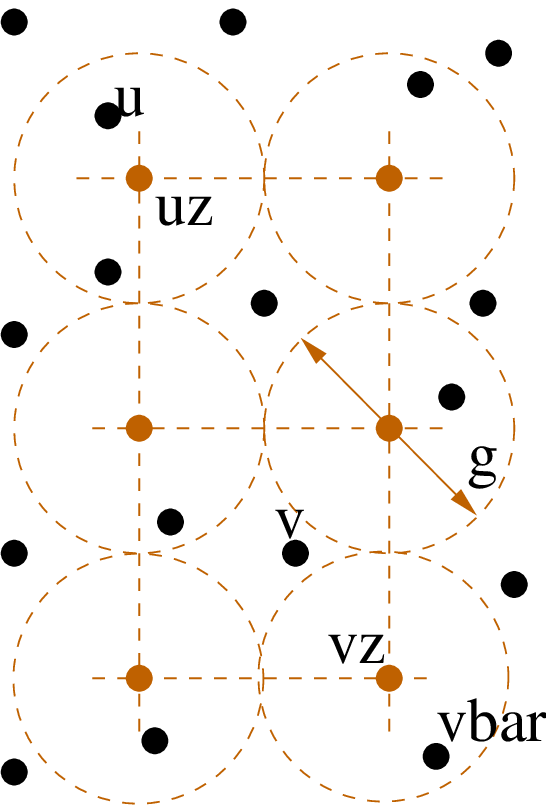}
\caption{\label{fig:dense_embedding_part2}Natural drawing of the $2$-D lattice (gray/brown) superimposed on
the $f'$ drawing of $\G$. Edges are not shown to prevent clutter. In this example, $u=\bar u$ but $v \neq \bar v$.}
\end{center}
\end{figure}
Now,
\begin{align*}
d_{\Z_d}(u_z,v_z) & = \|g(u_z)-g(v_z)\|_1\\
                  & \leq \sqrt{d}\|g(u_z)-g(v_z)\|,
\end{align*}
and
\begin{align*}
\|g(u_z)-g(v_z)\| & \leq \|g(u_z)-f'(\bar u)\|+\|f'(\bar u)-f'(u)\|+ \\
                  & \qquad \|f'(u)-f'(v)\|+ \|f'(v)-f'(\bar v)\|+\\
                  & \qquad \quad \|f'(\bar v)-g(v_z)\|.
\end{align*}
We know that $\|g(u_z)-f'(\bar u)\|  \leq \frac{1}{2} \leq \frac{\sqrt{d}}{2}$ since
$f'(\bar u) \in B(g(u_z),\frac{1}{2})$, and $\|f'(u)-f'(\bar u)\| <
\frac{3}{2}\sqrt{d}$ from~\eqref{eq:df-u-ubar}. Using these in the above, we get
\begin{align*}
& \|g(u_z)-g(v_z)\| \leq 4\sqrt{d}+d_{f'}(u,v),\\
\Rightarrow & d_{\Z_d}(u_z,v_z)\leq 4d+\frac{\sqrt{d}}{\gamma}d_{f}(u,v)
\end{align*}
which is the desired result.
\end{proof-theorem}

\bigskip

\medskip

\ifthenelse{\equal{\PaperORReport}{Report}}{

\begin{proof-proposition}{\ref{prop:examples}}
Proving that the graph in the first example has the required properties is straightforward and is therefore omitted. For the second example created from a 2-D lattice after removal of a certain fraction of its nodes, we first have to prove there is a giant connected component. First of all, $\alpha < \frac{1}{4(\beta+1)}$ implies there exists a positive integer $L$ such that $\alpha L^2 + \beta \le L-1$. Pick such a $L$ and superimpose a square grid on the natural drawing of the lattice $\Z_2$, with each side of the grid having length $L$; as shown in Figure~\ref{fig:gridonZ2}.
\begin{figure}
\begin{center}
\includegraphics[scale=0.4]{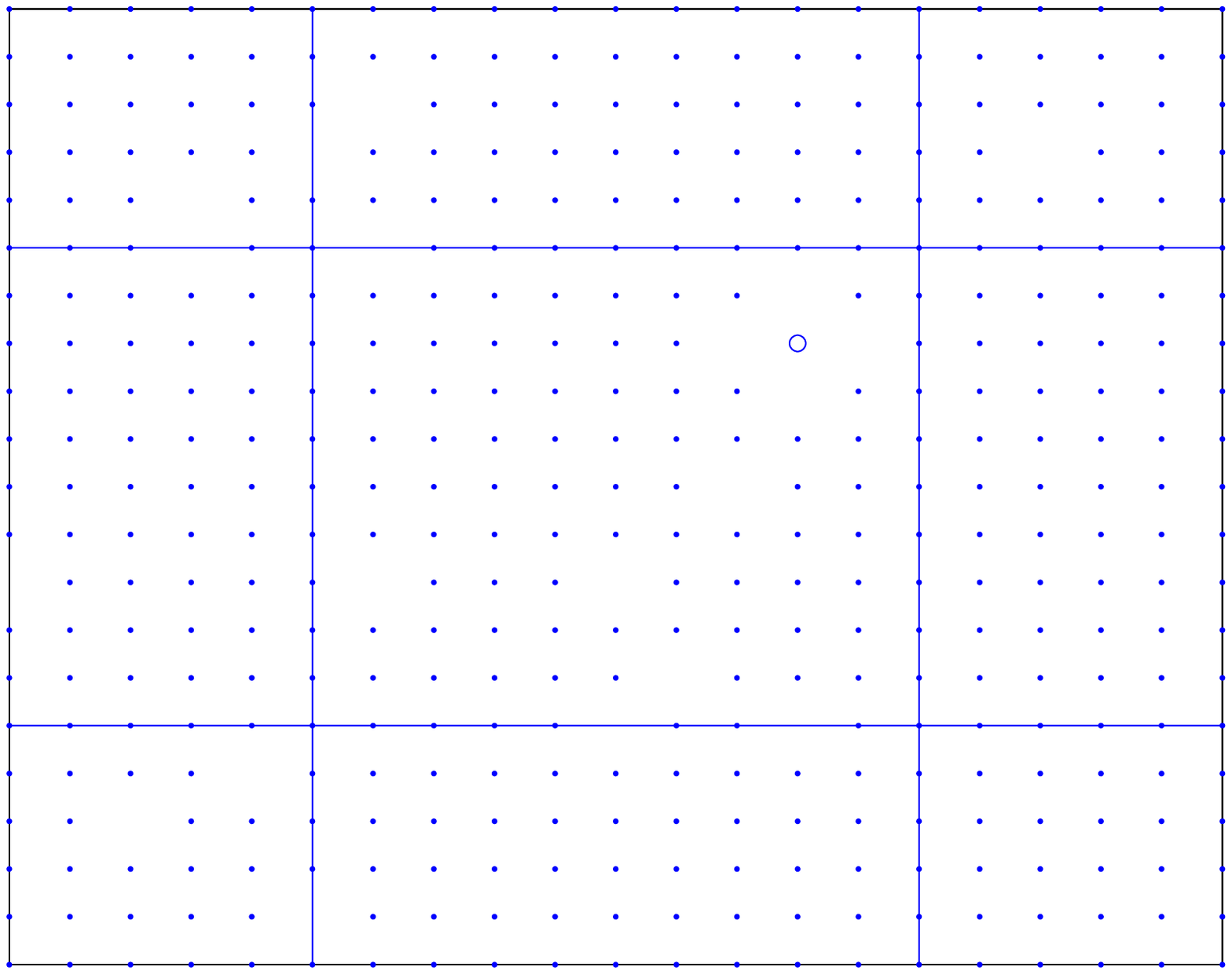}
\caption{2D lattice with certain fraction of nodes removed to simulate node failure. Nodes are shown as dots, except nodes that belong to disconnected components created due to failed nodes, which are shown as circles. Edges are not shown to avoid clutter. The superimposed grid has cell size equal to $L \times L$, cell boundaries being shown in solid lines.   }\label{fig:gridonZ2}
\end{center}
\end{figure} 
Pick an arbitrary cell $C$. The nodes in $C$ that have failed may divide the remaining nodes inside the cell into multiple connected components. Consider the largest connected subgraph formed by the remaining nodes inside the cell, and call it $\G_c = (\V_c, \E_c)$. The edges in $\E_c$ are incident on only those nodes that are entirely in $\V_c$.  There is at least one node on each of the four boundaries of $C$ that is part of $\G_c$. If not, all the $L$ nodes on at least one side must either fail or be removed from the largest component due to separation by the failed nodes. In both case, the number of nodes that must fail in a region of area $L^2$ is at least $L$, which means $\alpha L^2 + \beta \ge L$. However, this violates the condition that $\alpha L^2 + \beta \leq L-1$. Moreover, each of these boundary nodes are connected to one another by a path that lies entirely in $\G_c$. For, if not, there must be a ``fault line'' created by failed nodes that divides the cell into a left and a right (or, top and bottom) half, which requires the removal of $L$ nodes. By the argument present above, such a situation cannot occur. Therefore, the largest connected component in each cell contains at least one node from each of its four sides. 

\smallskip

Now consider two adjacent cells $C_1$ and $C_2$, sharing a common side of length $L$. Pick a node $b$ that belongs to $\G_{C_1}$ and lies on the common boundary between $C_1$ and $C_2$, which exists by the arguments above.  This node $b$ also belongs to $\G_{C_2}$, since otherwise this node has to be separated from the nodes in $\G_{C_2}$ by a ``fault line'' of failed/removed nodes, which was  impossible. Therefore, the graphs $\G_{C_1}$ and $\G_{C_2}$ are connected. Moreover, a conservative upper bound on the graphical distance between two arbitrary nodes in $\G_{C_1}$ and $\G_{C_2}$  is $8L$. Continuing this argument, we see that the largest connected component  in each cell is connected to those in adjacent cells, which proves that there is a giant connected component formed by the nodes after removal of the failed nodes from $\Z_2$.  We call this giant component $\G$. To prove that $\G$ is dense in $2$-D, first consider the natural drawing of $\G$ that is obtained from that of $\Z_2$ after removal of the failed nodes. The  largest uncovered diameter $\gamma$ in this natural drawing of $\G$ is trivially finite. To prove that $\rho > 0$, pick two arbitrary nodes $u$ and $v$ in $\G$. The minimum number of  adjacent cells of size $L \times L$ one has to go through to go from $u$ to $v$ is no greater than $\lceil \frac{\sqrt{2}d_f(u,v)}{L} \rceil $, where $d_f(\cdot,\cdot)$ is the Euclidean distance in the natural drawing of the lattice $\Z_2$. Therefore, the graphical distance between $u$ and $v$ satisfies  
\begin{align*}
d_{\G}(u,v) \leq 8L \lceil \frac{\sqrt{2}d_f(u,v)}{L} \rceil \leq 20 d_f(u,v).
\end{align*}
Lemma~\ref{lem:rho} now implies that $\rho > 0$, which proves denseness of $\G$. Sparseness of $\G$ follows trivially from the sparseness of $\Z_2$.
\end{proof-proposition}
}
{
}

\medskip

\begin{proof-lemma}{\ref{lem:lattice-dense-sparse}}
We only provide the proof that the $2$-dimensional lattice is not
sparse in $\R$ and is not dense in $\R^3$. The general case for
arbitrary dimensions is analogous.

To prove by contradiction the lack of denseness, assume that there
exists a dense drawing $f$ of $\Z_2$ in $\R^3$, with associated
$\gamma < \infty$ and $\rho > 0$.  Fix the origin of $\R^3$ at $f(u)$
for an arbitrary node $u$ in the lattice $\Z_2$. For an arbitrary
$D>0$, the volume of the sphere in $\R^3$ centered at the origin with
diameter $D$, denoted by $\ball^3 (0,D)$, is $\Omega(D^3)$. Therefore
the number of nodes of $\Z_2$ drawn inside $\ball^3(0,D)$ is
$\Omega((\frac{D}{\delta})^3) = \Omega(D^3) $. It is straightforward
to show that for any set of $n$ distinct nodes in the lattice $\Z_2$,
the maximum graphical distance between any two nodes in the set is
$\Omega(\sqrt{n})$. Therefore the maximum graphical distance between
the nodes in $\ball^3(0,D)$ is $\Omega(D^\frac{3}{2})$.

The maximum Euclidean distance between any two nodes drawn inside the
sphere $\ball^3(0,D)$ under the $f$-drawing is at most $D$, and since
$f$ is a dense drawing, it follows from Lemma~\ref{lem:rho} that for
every pair of nodes $u,v$ in $\Z_2$ such that
$f(u),f(v)\in\ball^3(0,D)$, we have $d_G(u,v) \leq aD+b$. Therefore,
the maximum graphical distance between pairs of nodes whose drawing
falls inside $\ball^3(0,D)$ is $\scr{O}(D)$, as well as
$\Omega(D^\frac{3}{2})$, which is a contradiction for sufficiently
large $D$. Hence no dense drawing of $\Z_2$ in $\R^3$ is possible.

\smallskip

To show $\Z_2$ is not sparse in $\R$, assume that there exists a civilized
drawing of $\Z_2$ in $\R$ with $s>0$ and $r<\infty$, where $r$ and $s$
are constants. Consider a subgraph $\Z_{2(n)}$ of $\Z_2$ that consists
of all nodes within a Euclidean distance $n$ from the origin. The
total number of nodes in this finite subgraph is $\Omega(n^2)$. The
length of the interval, $L$, in which the nodes of this subgraph are
located in the sparse $1$-d drawing of $\Z_2$ is clearly $L =
\Omega(sn^2)$. Since the maximum graphical distance between any two
nodes in the subgraph $\Z_{2(n)}$ is $n$ by construction, the maximum
connected range in the $1$-d drawing must be at least $r \geq
\frac{L}{n} = \Omega(sn)$. Since this must be true for every $n$, $r$
cannot be a finite constant. Thus, no civilized drawing of $\Z_2$ in
$\R$ exists.
\end{proof-lemma}

\ifthenelse{\equal{\PaperORReport}{Report}}{
\section{Relative position measurements from noisy range and angle measurements}\label{sec:unbiased-measurement}
Here we show how to convert noisy measurements of range and angle between a pair of nodes into a measurement of relative position between them such that the error in the measurement is zero mean. 

\medskip

Let the position of two nodes $u$ and $v$ lying on a plane (in a in a 2D Cartesian reference frame) be denoted by $x_u$ and $x_v$, respectively. Let the relative position between them (i.e., $x_u - x_v$)  be $[\Delta x, \Delta y]^T$, and the distance and angle between them (in the same reference frame) be $r$ and $\theta$. This implies $\Delta x = r \cos \theta$ and  $\Delta y = r \sin \theta$. Let $\widehat{r}$ and $\widehat \theta$ be noisy measurements of the distance and angle: $\widehat{r} = r + \delta r$, $\widehat{\theta} = \theta + \delta \theta$, where $\delta r$ and $\delta \theta$ are random errors in range and angle measurements. 

\medskip

The following is an unbiased measurement of $x_u- x_v$:
\begin{align}\label{eq:zeta}
  \zeta_{u,v} & = \begin{bmatrix}\widehat{\Delta x} \\ \widehat{\Delta y} \end{bmatrix}  =    
                  \frac{1}{\bar{c}}\begin{bmatrix}\widehat{r}\cos \widehat{\theta} \\                               \widehat{r}\sin \widehat{\theta} 
                  \end{bmatrix},    \text{   where } \bar{c} := \Exp[\cos (\delta \theta)],
\end{align}
under the assumption that (i) $\delta r$ and $\delta \theta$ are independent, (ii) $\delta r$ is zero mean, and (iii)  $\Exp[\sin \delta \theta ] = 0$.

\medskip

Before we show why the measurement error $\zeta_{u,v} - (x_u - x_v)$ is zero mean, we note that the measurement $\zeta_{u,v}$ can be obtained in practice from range and angle measurements and knowledge of $\bar{c}$.  The value of $\bar{c}$ can be obtained (estimated) a-priori while characterizing the AoA sensor used to gather angle measurements. For example, if $\delta \theta \sim N(0,\sigma^2)$, then $\bar{c} = e^{-0.5 \sigma^2}$ (which can be verified by symbolic integration in MATLAB). 

\medskip

To show that the measurement error $\zeta_{u,v}  - [\Delta x, \Delta y]^T$ is indeed zero mean, we expand the expression for $\widehat{\Delta x}$ to obtain
\begin{align}\label{eq:Deltax-hat}
 \widehat{\Delta x} = \frac{1}{\bar{c}}\left( \Delta x \cos (\delta \theta) - \Delta y \sin (\delta \theta) + \delta r \cos (\theta + \delta \theta) \right)
\end{align}
We decompose $\cos \delta \theta$ as
\begin{align}
\cos \delta \theta = \bar{c} + \tilde{c}_{\delta\theta},
\end{align}
where  $\Exp[\tilde{c}_{\delta\theta}] = 0$ by construction. From~\eqref{eq:Deltax-hat}, we get 
\begin{align}\label{eq:Deltax-hat-2}
 \widehat{\Delta x} & =  \Delta x  + \frac{1}{\bar{c}}\left(\tilde{c}_{\delta\theta} \Delta x - \Delta y \sin (\delta \theta) + \delta r \cos (\theta + \delta \theta) \right),
\end{align}
and similarly,
\begin{align}\label{eq:Deltay-hat}
\widehat{\Delta y} & = \Delta y + \frac{1}{\bar{c}}\left( \tilde{c}_{\delta \theta}\Delta y + \Delta x \sin \delta \theta + \delta r \sin (\theta + \delta \theta) \right),
\end{align}
which yields
\begin{align*}
  \zeta_{u,v} =
  \begin{bmatrix}
  \widehat{\Delta x} \\ \widehat {\Delta y}    
  \end{bmatrix} =
  \begin{bmatrix}
    \Delta x \\ \Delta y
  \end{bmatrix} +
  \begin{bmatrix}
    \epsilon_x \\ \epsilon_y
  \end{bmatrix}
\end{align*}
where 
\begin{align}\label{eq:epsilon}
\epsilon_x & := \frac{1}{\bar{c}}\left(\tilde{c}_{\delta\theta} \Delta x - \Delta y \sin \delta \theta + \delta r \cos (\theta + \delta \theta) \right) \\
  \epsilon_y & :=  \frac{1}{\bar{c}}\left( \tilde{c}_{\delta \theta}\Delta y + \Delta x \sin \delta \theta + \delta r \sin (\theta + \delta \theta) \right).
\end{align}
The random variables $\epsilon_x$ and $\epsilon_y$ are zero mean due to the assumptions on $\delta \theta$ and $\delta r$, which verifies the claim that the measurement error  $\zeta_{u,v} - (x_u - x_v)$  is indeed zero mean.

}
{
}

\section*{Acknowledgment}
The authors wish to thank Edmond Jonckheere for several useful
discussions. The first author wishes to thank Peter Doyle and J. L. Snell for the monograph~\cite{DoyleSnell}, from which he benefited immensely.

\end{document}